# List of Changes

(1) Page 4, paragraph 1, line 5: the different temperatures of the RTA are included.
(2) Page 5, Figure 2 caption: the meaning of W/O and W/ RTA is added.
(3) Page 8, Fig. 4: the values of spacing ($d$) and contact length ($z$) are indicated in both the figure and the caption.
(4) Page 3, paragraph 2, line 5: The acronym TLM is defined as "transmission line method".
(5) Page 23, the last three sentences: Additional discussion is provided to compare the photovoltaic results obtained in this study with those of similar HJ solar cells fabricated with silicon nanocrystals.
(6) Page 28: Two more references [51] and [52] are included to support newly added discussion.

# Response to Reviewer's Comments

(1) It is clear that the electrical properties of the Ge-NCs Tfs are good, but the interpretation of such properties should be related to the structural properties of the Ge-NC TFs (concentration and density of Ge-Ncs).
**Rsponse:** We totally agree that the correlation between structural and electrical properties is very important. But it is not easy to accurately determine the concentration and density of nanocrystals (intensive TEM work is needed). Therefore, we prefer not to include such discussion before conclusive data is obtained. The work is still under and we hope to report it later.

(2) The authors argue that electrical transport Ge-NCs Tfs is due to a thermally activated nearest neighbour hopping conduction. It is clear that such conduction mechanism is thermally activated (*ie* ln s $T^{-1}$ as shown in the article), but it is not sufficient. Hopping conduction is characterized by a temperature dependence following ln s $T^{-1/(1+d)}$ where d is the hopping space dimensionality.
**Response:** Thanks for mention this point. To our best knowledge, the conductivity-temperature relationship ln s $T^{-1/(1+d)}$ proposed by Mott [N. F. Mott, Philos. Mag. Vol. 19, pp. 835, 1969] is more suitable for describing the variable range hopping conductivity (VRH), where the hopping between state that are closer in energy (even if they are wider spaced) becomes more preferable than that between the nearest neighbors. It is also known that this conduction mechanism is more obvious at low temperatures. In the temperature range concerned in this study (above 200 K), it is very difficult to see such dependence, so that we do not adopt this model. Nevertheless, we are also interested to see if there is a transition to VRH region when temperature decreases and experiments are being conducted.

# Electrical Properties of Conductive Ge Nanocrystal Thin Films Fabricated by Low Temperature In-situ Growth

B Zhang[*], Y Yao, R Patterson, S Shrestha, M A Green and G Conibeer

ARC Photovoltaics Centre of Excellence, University of New South Wales, Sydney, New South Wales 2052, Australia

Email: bo.zhang@student.unsw.edu.au

**Abstract.** Thin films composed of Ge nanocrystals embedded in amorphous $SiO_2$ matrix (Ge-NCs TFs) were prepared using a low temperature in-situ growth method. Unexpected high p-type conductivity was observed in the intrinsic Ge-NCs TFs. Unintentional doping from shallow dopants was excluded as a candidate mechanism of hole generation. Instead, the p-type characteristic was attributed to surface state induced hole accumulation in NCs, and the hole conduction was found to be a thermally activated process involving charge hopping from one NC to its nearest neighbor. Theoretical analysis has shown that the density of surface states in Ge-NCs is sufficient to induce adequate holes for measured conductivity. The film conductivity can be improved significantly by post-growth rapid thermal annealing and this effect is explained by a simple thermodynamic model. The impact of impurities on the conduction properties was also studied. Neither compensation nor enhancement in conduction was observed in the Sb and Ga doped Ge-NCs TFs, respectively. This could be attributed to the fact that these impurities are no longer shallow dopants in NCs and are much less likely to be effectively activated. Finally, the photovoltaic effect of heterojunction diodes employing such Ge-NCs TFs was characterized in order to demonstrate its functionality in device implementation.

---

[*] Corresponding author. Tel.: + 61-2-93856782; Fax: +61-2-93855104; E-mail: bo.zhang@student.unsw.edu.au

1. **Introduction**

Group IV (Si and Ge) nanocrystals have attracted increased interest not only because of their potential application in nanoscale optoelectronic and photovoltaic devices, but also because of their full compatibility with large-scale integrated circuit fabrication [1-6]. Taking the advantage of its relatively low process temperature, Ge nanocrystals (Ge-NCs) are considered to be a good candidate for low cost manufacturing. In the past few decades, a lot of work has been done to develop synthesis methods and improve the structural and optical properties of Ge-NCs embedded in a $SiO_2$ matrix [5-17]. However, the implementation of this type of Ge-NCs thin films (Ge-NCs TFs) in semiconductor devices has rarely been demonstrated besides a few memory applications [6]. Such a situation can be partially attributed to the difficulty in making electrically conductive thin films, as well as the very limited understanding of the carrier generation and transportation mechanisms in these nanostructured thin films.

So far, only a few investigations have been done on the electrical conduction of Ge-NCs TFs. Inoue *et al.* studied very thin $SiO_2$ film (< 30 nm) containing 4 nm Ge-NCs [18]. They argued that the carriers selectively moved along the most conductive channel in the film. In a separate work, Fujii *et al.* reported different conductivity-temperature (σ-T) dependence for $SiO_2$ films containing Ge clusters (~ 2 nm) [19] and nanocrystallites (3.8 nm ~ 8.9 nm) [20]. The former composite exhibited a $\ln\sigma \propto T^{-1/4}$ relationship corresponding to the carrier conduction dominated by the variable range hopping process [21], while the latter one showed a $T^{-1/2}$ dependence on $\ln\sigma$ which could be explained by the percolation hopping theory. Later, Zhao *et al*. used a space charge limited conduction model to describe the I-V behaviour of Ge-implanted SiO$_2$ films [22]. Despite different conduction mechanisms, all these composites seem to have relatively low conductivities. This is a disadvantage for semiconductor devices requiring large current such as light emitting diodes and solar cells.

In this paper, the electrical conduction properties of Ge-NCs TFs prepared by magnetron co-sputtering and low temperature in situ NC growth was investigated. The thin films were found to exhibit p-type characteristics and high conductivities. The mechanism responsible for the generation of holes was first discussed. The current transport was then studied in detail using the transmission line method (TLM) and a temperature dependent measurement. Special attention was paid to the effect of post-deposition rapid thermal annealing (RTA) on the film conductivity. In addition, we also discussed the influence of impurity incorporation on the conduction properties. Lastly, we fabricated and characterized heterojunction devices to demonstrate the functionality of the Ge-NCs TFs as a semiconductor material.

2.  **Experimental**

Ge-NCs TFs of a thickness of 250 nm ~ 300 nm were deposited by co-sputtering using a RF magnetron sputtering apparatus. The composite target was made up of a circular fused quartz plate partially covered with high purity Ge strips (99.9999%). Six Ge strips, each with a centre angle of 12$^\text{o}$, were uniformly distributed on the quartz plate. The coverage of Ge was ~ 20% of the area of the composite target and X-ray Photoelectron Spectroscopy (XPS) suggested that this resulted in a Ge atomic concentration of ~ 35% in the thin films. The three types of Ge targets used in this study are listed in table 1. The sample names corresponding to respective targets are also included. During sputtering process the substrates were intentionally heated up to ~ 380 $^o$C for in-situ growth of Ge-NCs. Details of the co-sputtering and low temperature growth process are described in our previous work [17, 23]. Finally, RTA treatments at different temperatures, including 650 $^o$C, 700 $^o$C, 750 $^o$C, 800 $^o$C were carried out in nitrogen.

Table 1. Three types of Ge targets used in the experiments.

| Target Type | Target Resistivity (Ω-cm) | Dopant Concentration ($cm^{-3}$) | Sample Name |
|---|---|---|---|
| Undoped | > 30 | < $10^{14}$ | i:Ge-NCs |
| Ga-doped | 0.01~0.04 | $1 \times 10^{18}$ ~ $1 \times 10^{19}$ | Ga:Ge-NCs |
| Sb-doped | 0.005~0.02 | $1 \times 10^{17}$ ~ $1 \times 10^{18}$ | Sb:Ge-NCs |

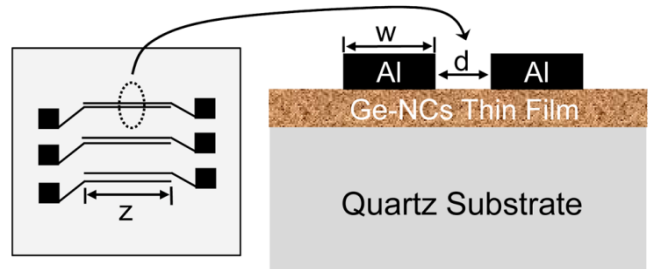

Figure 1. Schematic diagram of plane and cross-section views of the structures for electrical characterization.

The Ge-NCs TFs on quartz substrates were used for electrical characterizations. Metal contacts to the Ge-NCs TFs were formed by masked thermal evaporation of a ~ 700 nm thick Aluminium (Al) layer. The plane and cross-section views of the electrode pads are depicted in figure 1. There are three pairs of co-planar contacts with different spacings (*d*) ranging between 65 and 280 μm. The contact length (*z*) and width (*w*) are 1 cm and 330 μm, respectively. In this case, the measured current was most likely to flow through Ge-NCs TFs horizontally (along the X-direction). No sintering or forming gas anneal was performed after metallization.

## 3. Results and Discussion

3.1 *Structural analysis*

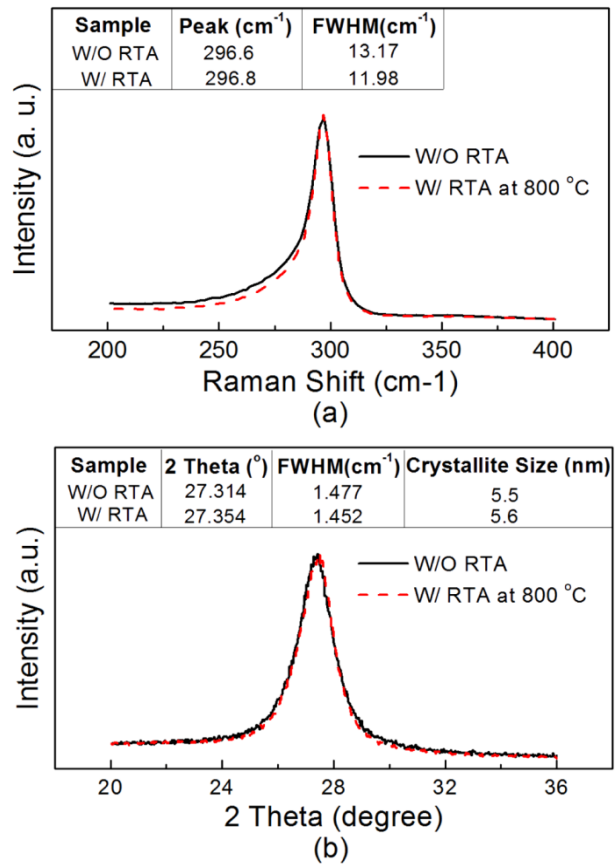

Figure 2. (a) Raman and (b) XRD spectra of Ge-NCs TFs without (labeled as "W/O") and with (labeled as "W/") RTA at 800 °C.

Essential structural properties were studied first to support subsequent discussion on electrical characteristics. The formation of Ge-NCs was confirmed by the Raman and X-ray diffraction (XRD) measurements as shown in figure 2. The figure gives the peak position, full width at half maximum (FWHM) and crystallite size of the Ge-NCs present. We can see that RTA did slightly improve the quality of the nanocrystals, as is evident from the narrower FWHM and the smaller hump in the low frequency tail in the Raman spectra of the RTA annealed film. Nevertheless, the difference in the spectra was on the whole very tiny. In the meantime, the XRD measurements exhibited almost identical spectra before and after RTA. Thus, the structural change of Ge-NCs and $SiO_2$ matrix was overall insignificant during RTA treatments. Particularly, it should be emphasized that the variation of the average crystallite size was negligible (~ 0.1 nm).

The size, distribution and crystal structure of the prepared Ge-NCs were further studied by high resolution transmission electron microscopy (HRTEM). This information will be needed in later theoretical calculations. The cross-sectional view in figure 3 confirms that quasi-spherical Ge-NCs were randomly dispersed in the amorphous silica matrix. The diameter of NCs was in a range of 3.8 to 8 nm, while the spacing between the NCs varied

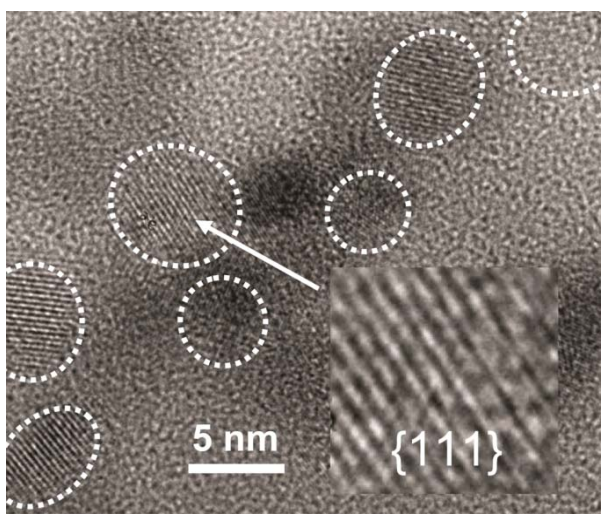

Figure 3. Cross-sectional HRTEM image of Ge-NCs TFs. The inset shows the lattice fringes

of the {111} planes of diamond structure Ge.

from 1.5 to 4.5 nm. Clear fringes observed in the lattice image were indicative of good crystallinity. In addition, the fringe spacing of ~ 0.33 nm and the angle between the crossed fringes of 70.5 ° were identified as the {111} planes of diamond-structure Ge.

*3.2 Electrical Properties*

*3.2.1 Conductivity of i:Ge-NCs TFs*

The current conduction characteristics of i:Ge-NCs TFs were measured using HP4140B pA Meter/DC voltage source with tri-axial wires to eliminate noise. The I-V curves of samples with different RTA treatments are displayed in figure 4(a). Straight lines were obtained for all samples except the one with RTA at 800 °C (in the inset), in which a slight non-linear effect emerged at low voltages. In this case, the conductivities were extracted from the linear regions at higher voltages (indicated by circles in the figure).

On the whole, fairly good ohmic contacts were formed between Al electrodes and Ge-NCs TFs, ensuring that the contribution from contacts to the conduction properties was negligible and film sheet resistance $R_S$ and contact resistance $R_C$ can be precisely extracted by TLM [24]. The TLM measurement was implemented with the lateral multiple-contact structure as shown in figure 1, in which the contact length (1 cm) was much longer than the "transfer length" (<0.1 μm). The conductivity of the as-deposited film was ~ 2.6 x $10^{-4}$ S/cm, which was markedly higher than previous results [18-20, 22]. Interestingly, the film conductivity could be further improved by post-deposition RTA. The film with 800 °C RTA reached a conductivity value as high as ~ 0.9 S/cm. This result was very consistent with Hall effect measurement which determined a carrier concentration of 3.2 x $10^{18}$ $cm^{-3}$ and a mobility of 2.25 $cm^2$/V·S. We also show the dependence of $R_S$ and $R_C$ on RTA temperature in figure 4(b). It is observed that both $R_S$ and $R_C$ were reduced by orders of magnitude after the RTA treatments. However, $R_C$ was always two orders of magnitude smaller than $R_S$, which could explain the observed ohmic contact behaviour in our samples.

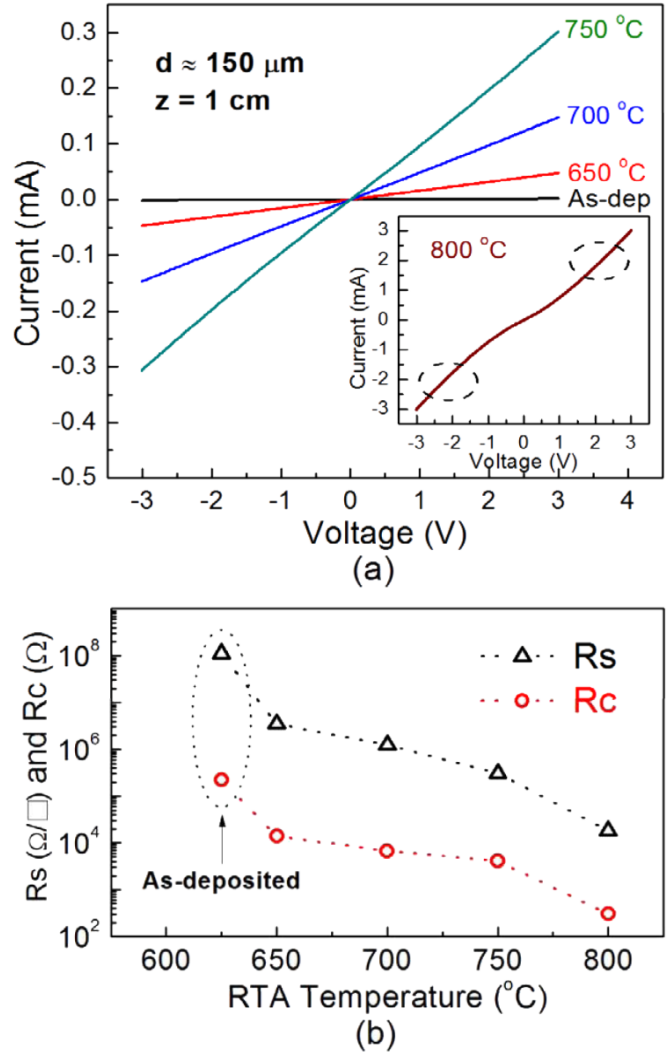

Figure 4. (a) I-V characteristics of i:Ge-NCs TFs with different RTA treatments. The contact length $z$ = 1 cm and the spacing between contacts d ≈ 150 μm. The as-deposited film is also included for reference and labeled "As-dep". The inset shows the I-V curve of the sample with 800 °C RTA. (b) The $R_S$ and $R_C$ of the same group of samples. The resistance of as-deposited film is indicated by the dotted circle.

*3.2.2 Hole Generation Mechanism*

The relatively high conductivity and carrier concentration in the i:Ge-NCs TFs is quite surprising, taking into consideration that they were not intentionally doped. The Hall effect and thermoelectric power measurements showed that holes were responsible for this charge transport. Hence, it is necessary to first understand the mechanism accounting for the carrier generation before other effects can be discussed.

One of the possible origins of holes is the unintentional incorporation of impurities into the films during sputtering or post-annealing process. Assuming acceptor impurities were the dominating contributor of charge carriers, we can correlate the generated hole concentration ($p$) with the acceptor concentration ($N_A$) and activation energy needed for the generation process ($E_G$). The relation can be expressed as following:

$$p \approx \sqrt{\frac{N_A N_V}{2}} \exp(-\frac{E_G}{2kT}) \qquad (1)$$

where $N_V = 6.0 \times 10^{18}$ cm$^{-3}$ is the effective density of states in the valence band of Ge, and $k$ is Boltzmann's constant. For the aforementioned sample with 800 °C RTA, we have $p \approx 3.2 \times 10^{18}$ cm$^{-3}$ and $E_G \approx 0.07$ eV (the derivation will be given later when discussing the temperature dependent I-V). The result of the calculation suggests that a minimum $N_A$ of ~ 5.15 x 10$^{19}$ cm$^{-3}$ is needed to achieve the measured hole concentration. Actually, this value is much lower than that required in a real case, since we have not considered the activation rate of dopants which is usually very low. If any impurity satisfying the above requirement is present in the i:Ge-NCs TFs, it should be detectable by X-ray photoelectron spectroscopy (XPS), as the lower limit of XPS resolution is ~ 1 x 10$^{19}$ cm$^{-3}$. However, the XPS measurement displayed no peaks other than those arising from Ge, Si and O. This observation suggests that the large amount of holes in the i:Ge-NCs TFs were very unlikely to result from impurity doping process. In fact, introducing majority carriers into NCs by conventional doping strategy is a challenge because of the fundamental physical phenomenon occurring in NCs, such as self-purification, oxidizing at the surface, high ionization energies and compensation by charge traps [25-28].

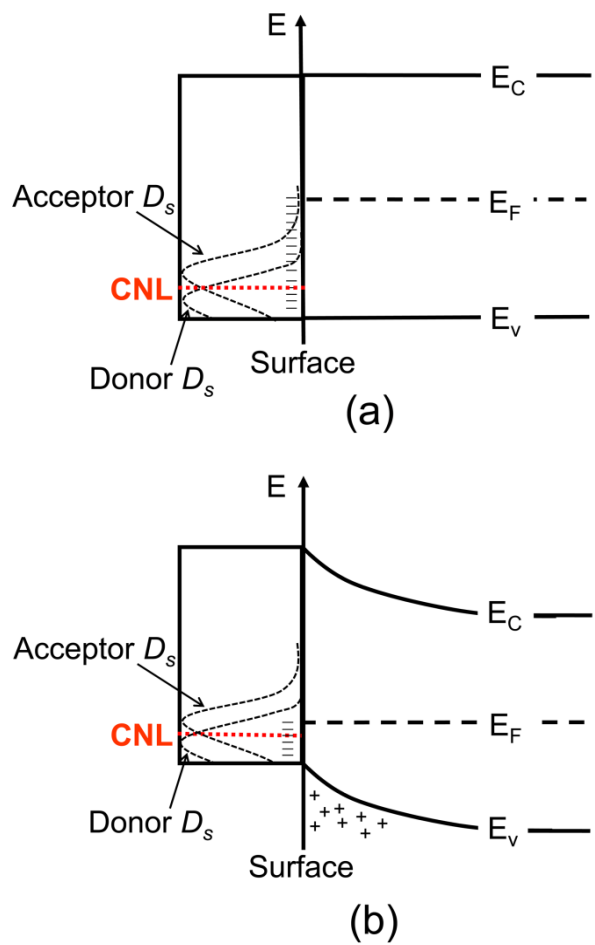

Figure 5. (a) Negative surface charges in bulk Ge, for the case that $E_F$ is well above the CNL and no surface band bending occurs; (b) band bending at the surface of bulk Ge, which results in hole accumulation. Gaussian energy distribution is assumed for acceptor and donor surface states (Ds) to facilitate the visualization of asymmetrical distribution of the effective density of surface states. The (-) and (+) denote the negative surface charges induced by filled acceptor-like states and the positive charge due to surface band bending, respectively.

**10 / 28**

After excluding the unintentional doping mechanism as the cause of the high film conductivity, we must turn to another possibility, the surface states. Extensive studies have revealed that these surface states of Ge are deep-acceptor-like states and are usually related to the unpassivated dangling bonds [29, 30]. It is also found that the surface states would give rise to an accumulation of holes at the Ge surface. This effect is shown in figure 5. Due to the asymmetrical distribution of the effective density of surface states [31], the charge neutrality level (CNL) in Ge lies very close to the valence band and well below the Fermi level ($E_F$) of intrinsic Ge, as illustrated in figure 5(a). A fixed negative charge is then built up at the surface as a result of the occupation of dominant unpassivated acceptor-like surface states. As a consequence, the energy bands near the surface tend to bend up and attract extra holes in the bulk region until the positive charge originating from accumulated holes and donor surface states balances the negative acceptor states charge, reestablishing charge neutrality in the material and pinning the $E_F$ near the CNL (figure 5(b)).

Very recently, such an effect has been employed to account for the hole generation and conduction observed in Ge nanowire and Ge/Si Core-Shell nanowire [32-34]. In NCs, the total surface area is much larger than that in bulk material and nanowire, so that it is reasonable to expect a more noticeable generation of holes. Furthermore, we also need to be aware of the size effect. Since NCs' size is very small compared with the typical width of the space charge region in bulk semiconductor of ~ 100 nm, the $E_F$ in NCs is likely to be shifted down across the entire crystal instead of band bending at the surface. This effect can also greatly enhance the efficiency of hole generation in Ge-NCs.

*3.2.3 Carrier Transport Mechanism*

In order to further explore the transport mechanism of surface state induced holes, temperature dependent film conductivities were measured in this study. The temperature range of the measurement is between 210 K and 320 K. The experimental setup is shown in figure 6(a). The samples were placed on a cooling/heating stage, the temperature of which was computer controlled. A temperature sensor on a piece of quartz with the same thickness as the sample substrate was mounted in close proximity to the samples in order to minimize the difference between set temperature and sample temperature.

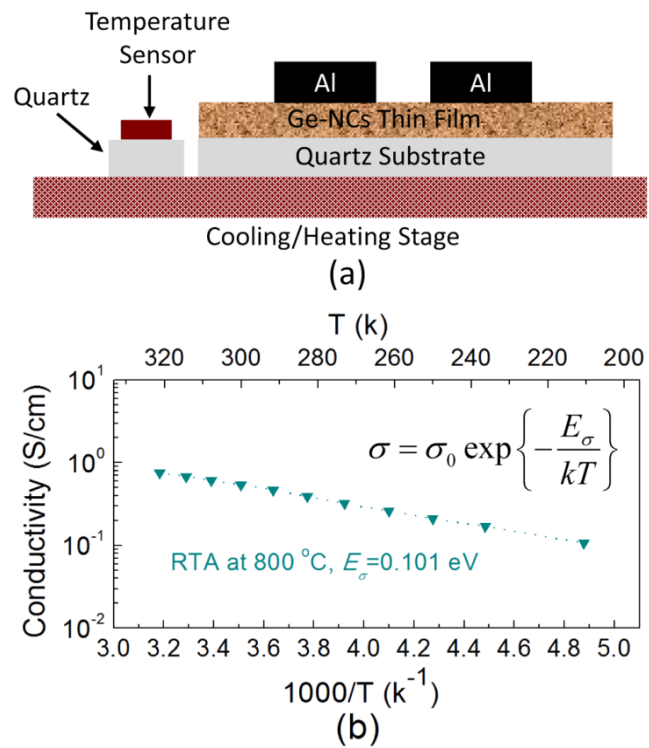

Figure 6. (a) Schematic diagram in cross section of the experimental setup for temperature dependent I-V measurement; (b) Electrical conductivities as a function of $1/T$ for the sample with 800 °C RTA. General expression for conductivity is also shown and $E_\sigma$ is the corresponding activation energy for the thermally activated conductivity.

The conductivities measured from the i:Ge-NCs TF with 800 °C RTA are plotted as a function of $T^{-1}$ (Arrhenius plot) in figure 6(b). The relationship is clearly linear, which is quite different from the conductivity-temperature dependence captured by Fujii *et al.* [19, 20]. Same behaviour was also observed in the as-deposited film and the films with different RTA temperatures. This strongly indicates a thermally activated nearest neighbor hopping conduction. In a nanocomposite system containing many semiconductor NCs, such as our Ge-NCs TFs, charge carriers have to be first excited to a transport energy level ($E_T$) and then hop to neighbor empty states by surmounting the electrostatic charging energy of NCs ($E_C$) [20]. Hence, the activation energy of conductivity $E_\sigma = E_G + E_C$, where $E_G$ is the energy associated with carrier generation. Accordingly, we are able to express the thermal-equilibrium concentration of holes $N_h$ at the transport energy level $E_T$ as:

$$N_h = N_T \exp(-\frac{E_\sigma - E_C}{kT}) \qquad (2)$$

where $N_T$ is the effective density of surface states at $E_T$. We can roughly evaluate $N_T = C_{surface} / t_{shell}$, where $C_{surface} = 10^{13} \sim 10^{14}$ cm$^{-2}$ is the total surface state concentration of Ge [29] and $t_{shell} \approx 1$ nm is the approximate thickness of the shell layer where surface states can exist [32, 35-37]. As can be verified from the equation embedded in figure 6(b), $E_\sigma$ of the 800 °C annealed film can be determined to be ~ 0.101 eV by taking the slope of the straight line. The $E_C$ is given as [20]:

$$E_C = \frac{e^2}{4\pi\varepsilon\varepsilon_0} \frac{1}{d} \frac{(s/d)}{(0.5 + s/d)} \qquad (3)$$

where $\varepsilon \approx 3.9$ is the dielectric constant of SiO$_2$ matrix, $\varepsilon_0$ is the permittivity in vacuum, and $s \approx 3$ nm and $d \approx 6$ nm are average NC spacing and diameter estimated from the HRTEM image (figure 3), respectively. From equation (2), we can then obtain a range of $N_h$ that is between 6.7 x $10^{18}$ and 6.7 x $10^{19}$ cm$^{-3}$. This theoretical value is in good agreement with the results of Hall effect measurements that showed a hole concentration of ~ 3.2 x $10^{18}$ cm$^{-3}$, indicating that surface trap states were present in sufficient concentration to provide observed electrical conductivities. In addition, the slightly larger theoretical value may result from neglecting carrier trapping by matrix defects in the calculation, and meanwhile the $E_C$ or $N_T$ may also be overestimated.

3.2.4 *Effect of RTA on Conductivity*

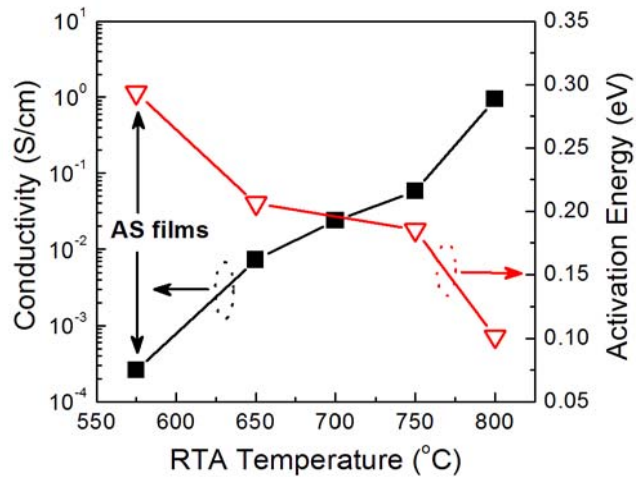

Figure 7. Conductivities and activation energies of i:Ge-NCs TFs as a function of RTA temperatures. The as-deposited films are also included and labeled as "AS films".

As can be seen from figure 7, the post-deposition RTA treatments significantly increased the conductivity of Ge-NCs TFs. At the same time, the activation energies for conductivity ($E_\sigma$) gradually decreased with increased RTA temperatures. It seems that the improvement of film conductivity was likely to be related to the reduction of $E_\sigma$. From previous discussion we know that $E_\sigma$ consists of $E_G$ and $E_C$. $E_C$ was considered to stay constant during RTA treatments, since the size and distribution of Ge-NCs did not appear to be influenced by it. Hence, $E_G$ must decrease correspondingly and this probably contributed to the enhancement of film conductivity.

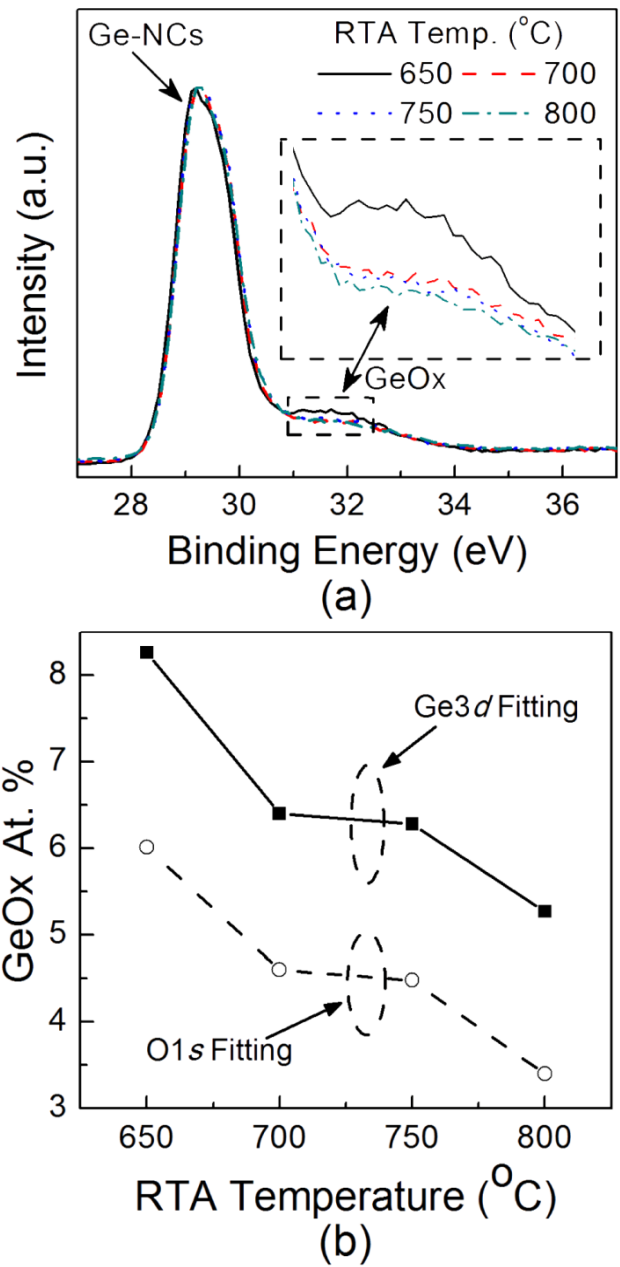

Figure 8. (a) XPS spectra of Ge-NCs TFs with RTA at different temperatures. The humps indicated by dashed circle and enlarged for clarity are related to the Ge suboxide ($GeO_x$)

component; (b) The GeO$_x$ at.% as a function of RTA temperature from both Ge3$d$ fitting and O1$s$ fitting.

In theory, the reduction of $E_G$ can be explained in terms of a $E_F$ shift towards the transport path level ($E_T$). It has been illustrated earlier that surface states rather than impurity doping gave rise to the hole accumulation in Ge-NCs and thereby the $E_F$ shift. Consequently, we conjecture that high temperature annealing process might have modified the surface structure of Ge-NCs. XPS spectra around the core level Ge 3$d$ have been employed to examine the effect of RTA on the surface chemical bonding structure. The spectra of samples with different RTA temperatures are displayed in figure 8(a). Besides the sharp peaks that are related to Ge-NCs (at binding energy of ~ 29.3 eV), all the spectra exhibit a small hump between 30.5 eV and 33.5 eV (indicated by dotted rectangle), which can be ascribed to a mixture of Ge substoichiometric oxides (GeO$_x$ where x<2). We expect that these GeO$_x$ components mainly presented in an interface layer between Ge-NCs and SiO$_2$ matrix as shown in figure 9 [37]. The formation of GeO$_x$ can be explained by the fact that the surface atoms of Ge-NCs tended to be terminated by oxygen atoms under an oxygen rich deposition environment. It is further noticed that these GeO$_x$ related humps slightly decreased with increased RTA temperature (shown in the enlarged rectangle). In order to quantitatively show this variation, the spectra were numerically fitted by multiple Gaussian functions. As shown in figure 8(b), curve fitting of Ge3$d$ spectra indicates that the atomic ratio of GeO$_x$ (GeO$_x$ at.%) gradually decreased from 8.26% to 5.27% when the RTA temperature increased from 650 °C to 800 °C. Also included in this figure is the evolution of GeO$_x$ at.% estimated from the curve fitting of O1$s$ spectra, which shows a little difference in absolute values but nearly same variation tendency. Since Ge-O bonds are thermally unstable and dissociate at elevated temperatures higher than 500 °C [38], above observed loss of GeO$_x$ component can be attributed to the decomposition of surface Ge-O bonds during high temperature post-annealing process. As a result of the bond breaking, extra surface dangling bonds (DBs) were present in the samples (figure 9), which enhanced the hole accumulation effect in Ge-NCs and thereby improved film conductivity. This simple thermodynamic model also implies that higher RTA temperature can provide a better film conductivity, by causing a more severe decomposition of surface Ge-O bonds.

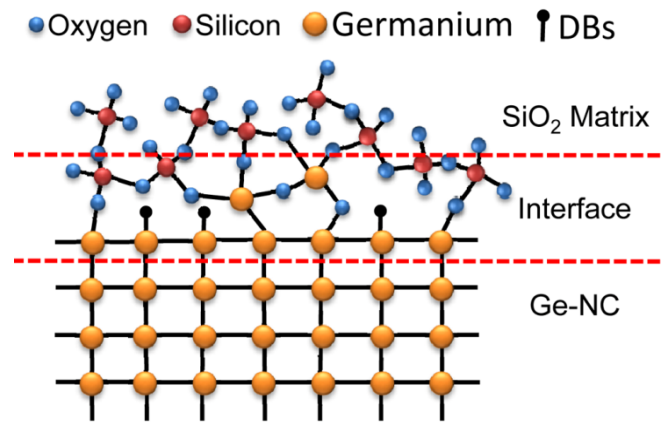

Figure 9. Schematic diagram of the GeO$_x$ components and the dangling bonds (DBs) within the interface layer between Ge-NCs and SiO$_2$ matrix.

Additionally, Fourier Transform Infrared Spectroscopy (FT-IR) was used to investigate the structure of SiO$_2$ matrix for different RTA temperatures. All measured spectra in figure 10 illustrate similar characteristic peaks of SiO$_2$ matrix [39]. However, when RTA temperature increased from 650 °C to 800 °C, we notice that the center of the Si-O-Si asymmetric stretching band gradually shifted from 1044 to 1064 cm$^{-1}$, as indicated by the arrows in the inset. Simultaneously, the FWHM of the absorption bands also decreased with increased annealing temperature. It is known that such blue shift of IR absorption band and narrowing of bandwidth are usually attributed to the stoichiometric transition from suboxide (SiO$_x$, 1< x <2) to silicon dioxide [40, 41]. In our case, the additional oxygen atoms required for this procedure to occur could be partially supplied from the aforementioned dissociation of Ge-O bonds at the surface of Ge-NCs. To some extent, this stoichiometric change indicated a reduction of oxygen-deficiency-related defects during RTA treatment, and thus more hole charges could participate in the electrical conduction rather than being trapped by defect centers.

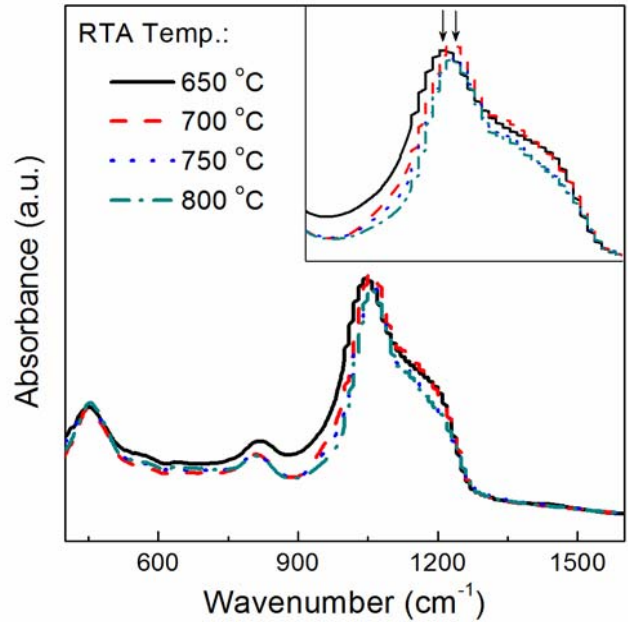

Figure 10. FT-IR spectra of Ge-NCs TFs for different RTA temperatures, from 600 $^{\circ}$C to 800 $^{\circ}$C. The inset is the enlarged image of the spectra in the 850 ~ 1300 cm$^{-1}$ wavenumber range.

Besides carrier density, carrier mobility is another primary factor that affects the film conducitivity, as stated in the definition of conductivity $\sigma = en\mu$, where $e$ is electron charge, $n$ is free carrier density and $\mu$ is carrier mobility. The effect of post-deposition RTA on the hole mobility of Ge-NCs TFs is still unclear in this study. This is because, despite of the increased carrier density and conductivity of the films, the Hall effect measurements did not have sufficient sensitivity to provide accurate results. Taking into consideration the similar carrier scattering between samples, the mobility was actually not likely to vary much over the temperature range of RTA. In future work, we will try alternative methods to measure the small mobility, such as photoconductivity measurement with a time-of-flight setup.

3.2.5 *The Impact of Impurities on Film Conduction Properties*

In this section, the influence of Sb and Ga impurities on the electrical properties of Ge-NCs TFs will be reported. As described in the experimental part, highly doped Ge targets were used to introduce dopant impurities. Resistivity data allows the target doping concentration to be estimated and it is roughly in the range of 1 x $10^{17}$ ~ 1 x $10^{18}$ cm$^{-3}$ for Sb and 1 x $10^{18}$ cm$^{-3}$ ~ 1 x $10^{19}$ cm$^{-3}$ for Ga [42]. The same electrical characterization techniques as used for i:Ge-NCs TFs have been repeated to study the Ga:Ge-NCs TF and Sb:Ge-NCs TF.

It is seen from figure 11 that the Ga/Sb:Ge-NCs TFs basically had similar conductivities and activation energies as the intrinsic TFs throughout the entire range of RTA temperatures. At the same time, we found that the Sb:Ge-NCs TFs exhibited a p-type electrical conduction rather than n-type conduction. Thus we can deduce that only a slight amount of dopants were activated in those doped samples, and the "macroscopic" doping compensation or enhancement effect was insignificant. On the contrary, recent works focusing on the doping of

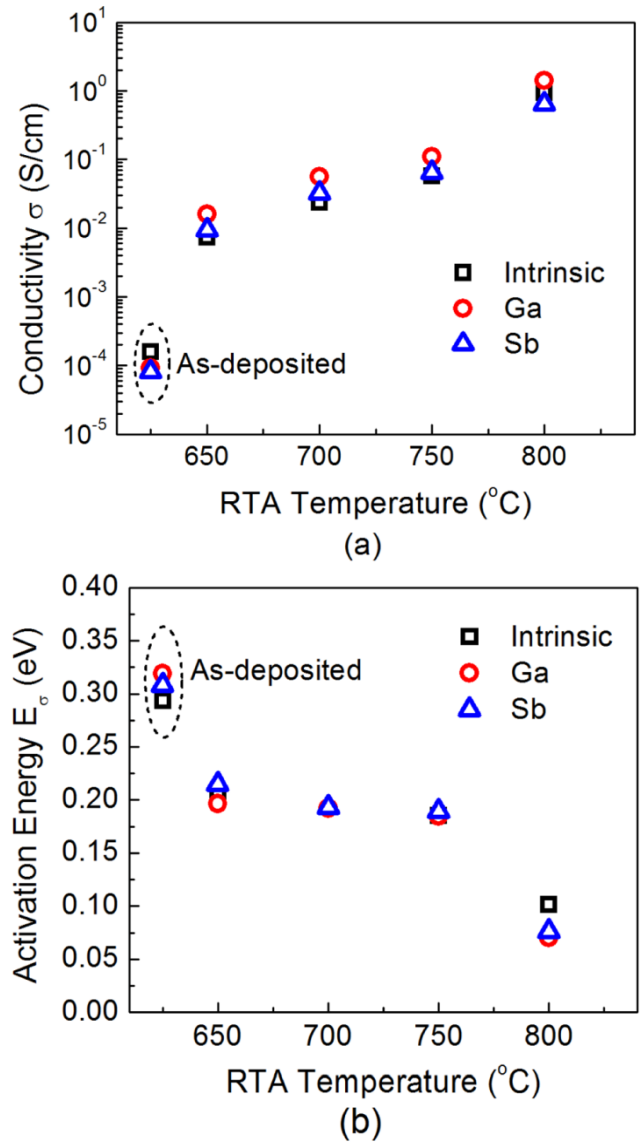

Figure 11. The evolution of (a) film conductivities and (b) activation energies with RTA temperature for the films sputtered from intrinsic and doped targets. Square: intrinsic target; Circle: Ga doped targets; Triangle: Sb doped targets.

Ge nanowires, especially for n-type materials, have shown efficacious incorporation and activation of dopants by forming a heavily doped shell surrounding an undoped (or very lightly doped) core [33, 43-45]. The dopants incorporation pathway was predominately via the conformal surface growth of the shell layer and thereby the concentration of dopants was determined by the rates of dissociative chemisorptions of precursors. It is worthy of attention that a doping level higher than $1 \times 10^{19}$ cm$^{-3}$ (the highest value was $\sim 2 \times 10^{20}$ cm$^{-3}$) was obtained in the nanowires that displayed a distinct n-type behavior. By contrast, our Ge-NCs TFs were more lightly doped (up to $1 \times 10^{18}$ cm$^{-3}$ for Sb) and this disparity may partly explain the lack of compensation effects. More importantly, we have not considered the more significant quantum confinement effect in NCs. As a result of relatively large shell thickness (10 ~ 20 nm), the spatial confinement of electrons (holes) along the direction perpendicular to the surface is very weak in above mentioned nanowires. Hence, the donors or acceptors incorporated into the shells can still be considered as shallow and can be activated effectively. The situation is quite different for Ge-NCs in the size regime we are concerned within this work. The large values of the binding energy resulted from the strong physical confinement of the electrons (holes) within NCs suggest that the dopants probably cannot be considered as shallow in small crystallites [46]. Simultaneously, the ionization energy may also be increased by the discretization of the continuum of conduction band states in NCs [27]. Therefore, NCs should have a much lower dopant activation rate than nanowires and this further prevents the doping effect in our samples. These results actually agree very well with earlier work on Si-NCs in SiO$_2$ matrix, which also reported an inefficient incorporation and activation of nominally "shallow" dopants into Si-NCs TFs [47]. As we have seen in this paper, Ge-NCs inherently possess a strong p-type characteristic, so it will be even more of a challenge to produce n-type Ge-NCs. A few alternative strategies are currently being investigated within our group. One of the possibilities is to dope the SiO$_2$ matrix surrounding the NCs instead of directly doping the NCs. This method has similar concept as the "modulation doping" technology widely used in the preparation of compound semiconductor devices [48].

Although direct physical evidence of dopant location was not provided in this study because of difficulties in measurements, a segregation of dopant impurities to the NCs' surface was expected to occur during the thermal annealing process involved in the preparation of the TFs (either in-situ substrate annealing or post-deposition RTA). The presence of impurity atoms in the surface region may change the surface structure and chemical bonding states, modifying the distribution of defect energy levels within the band gap. This hypothesis may account for the difference of activation energies in the undoped and doped Ge-NCs TFs with RTA at 800 $^{o}$C (figure 11(b)). However, we should also be aware that such difference was not observed in the samples post-annealed at lower temperatures. This could suggest to some extent that certain thermal energy is needed for the incorporated impurities to cause a substantial change of the surface states.

3.3 *Heterojunction Formation using Ge-NCs TFs*

Heterojunction (HJ) devices employing Ge-NCs TFs on lightly doped n-type crystalline silicon substrates with impurity concentration of ~ 1 x $10^{15}$ cm$^{-3}$ (n:c-Si) were fabricated to evaluate the integratability of the nanostructured thin film and investigate the design parameters required for its application in photovoltaic devices. The schematic diagram of the HJ device is illustrated in figure 12. The total area of the device is 1 cm$^2$. The thickness of the Ge-NCs layer was about 250 nm. The Ge-NCs TFs were post-annealed by RTA at 800 $^{o}$C in order to achieve smaller film resistivity and higher carrier concentration. The Al front fingers and rear contact were deposited by thermal evaporation. No passivation or sintering process was performed on the devices.

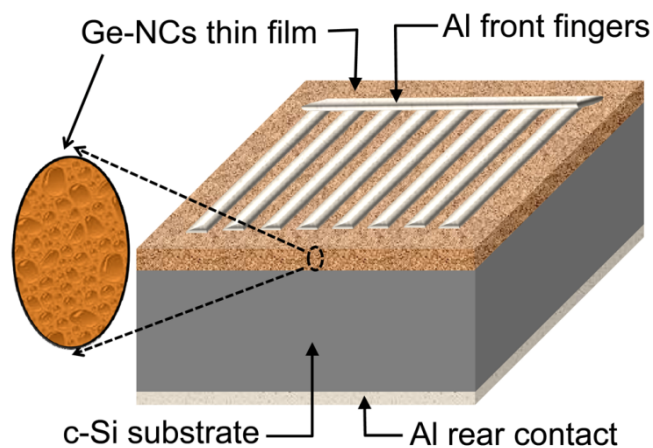

Figure 12. Schematic diagram of a Ge-NCs/n:c-Si HJ diode.

Figure 13 shows the dark I-V curves and photovoltaic properties of a typical i:Ge-NCs/n:c-Si HJ device at room temperature. The device shows good current rectification of three orders of magnitude at ± 1.5 V. Since both front and rear electrodes were ohmic contacts, the rectification effect should be attributed to the junction. The photovoltaic properties of the device were also evaluated by the quasi-steady-state open circuit voltage (Suns-$V_{oc}$) measurement [49]. The advantage of this method is that it can eliminate the adverse effects of series resistance on $V_{oc}$, which are significant in our samples. Apparent photovoltage was detected from the illuminated device and the 1-sun $V_{oc}$ was found to be ~ 314 mV. The best fitting to the experimental data using a two diode model predicts an effective ideality factor of ~ 1.01 throughout entire injection range, which is indicative of a dominating bulk and surface recombination in the heterojunction devices [50]. These preliminary results are encouraging as a starting point for using Ge-NCs in photovoltaic applications, though they are still not comparable with similar HJ solar cells employing Si-NCs which exhibited $V_{oc}$ higher than 500 mV [51, 52]. Nevertheless, it should be aware that those Si-NCs thin films were very heavily doped and high temperature annealed, so that better p-n junctions and higher free carrier concentration could be expected. In other words, this would mean that there is still a lot of space to improve the Ge-NCs thin films.

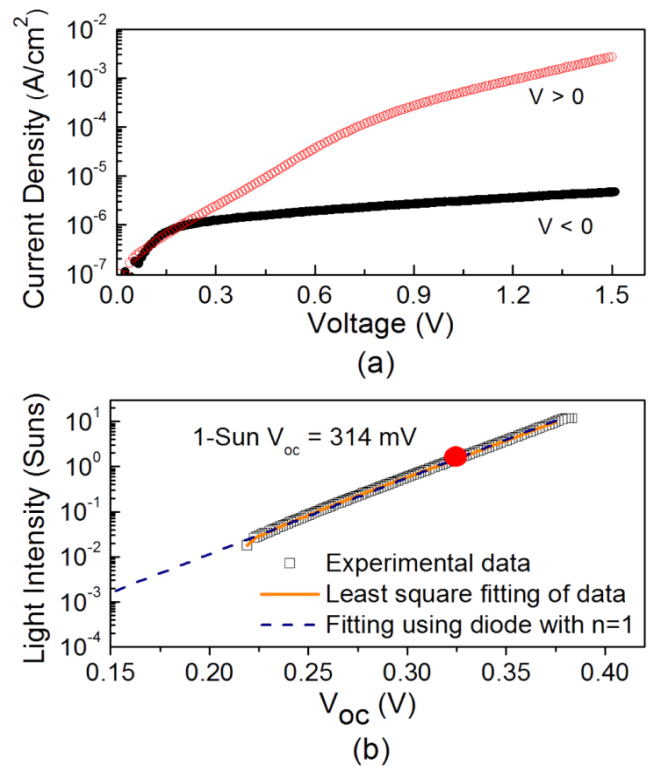

Figure 13. (a) The semilogarithmic plot of dark I-V curves of a typical i:Ge-NCs/n:c-Si HJ device in both polarities at room temperature. (b) The measured suns-$V_{oc}$ characteristics of the HJ device. The red solid circle indicates the 1-sun point.

The two effects mentioned above, current rectification and photovoltage, are indeed good evidence for the formation of p-n junctions in the i:Ge-NCs/n:c-Si HJ devices. In contrast, HJ devices with i:Ge-NCs on p-type silicon substrates exhibited neither current rectification nor photovoltaic effects, probably due to the lack of a p-n junction. Thus it can be concluded that the i:Ge-NCs TFs behave like a p-type doped semiconductor material. Moreover, same phenomenon has been observed in the HJ devices employing Ga:Ge-NCs and Sb:Ge-NCs, which is consistent with our previous arguments that the doped Ge-NCs TFs also have a p-type characteristic. More quantitative analysis of these devices is beyond the scope of this paper and will be reported elsewhere.

4. **Conclusion**

We have demonstrated in this paper that highly conductive p-type Ge-NCs TFs can be prepared by in-situ low temperature growth technique and subsequent RTA process. By excluding the possibility of unintentional doping from impurities, the inherent p-type characteristic of Ge-NCs TFs was ascribed to the hole accumulation in NCs caused by the dominating acceptor-like surface states. Temperature dependent measurement revealed a $\ln\sigma \propto T^{-1}$ relationship, suggesting a thermally activated nearest hopping conduction mechanism in these films. The carrier transport was considered to occur at the surface state energy level and a theoretical calculation predicts that the density of surface states of Ge can provide sufficient free holes to explain the observed conductivity. RTA treatments further increased the film conductivity without changing much of the structural properties of NCs. This improvement was tentatively attributed to the modification of surface structure of NCs and reduction of oxygen-deficiency-related defects in the $SiO_2$ matrix. The effect of incorporating moderate amount of Ga and Sb dopants was also investigated. The doped films exhibited similar conduction properties as the intrinsic films, which means the films were still dominated by surface state induced hole conduction and the dopants were not effectively activated. This is not surprising if one realizes the screening of shallow dopants in NCs due to the increase of binding energy and ionization energy. This effect together with the inherent hole generation effect actually make it very challenging to produce n-type Ge-NCs TFs. Finally, the Ge-NCs TFs were used to fabricate heterojunction diodes. Clear photovoltaic effect was observed from the devices, which demonstrated that such conductive Ge-NCs TF is a promising candidate material for low cost nanodevices.

5. **Acknowledgements**

This work was supported by the Australian Research Council (ARC) via its Centers of Excellence scheme and by the Global Climate and Energy Project (GCEP) administered by Stanford University. Bo Zhang thanks the Asia-Pacific Partnership on Clean Development and Climate, and IDP Education Australia for supporting his study in Australia.

**References**
[1] Pavesi L, Dal Negro L, Mazzoleni C, Franzo G and Priolo F 2000 *Nature* **408** 440
[2] Iacona F, Franzò G and Spinella C 2000 *J. Appl. Phys*. **87** 1295
[3] Ohba R, Sugiyama N, Uchida K, Koga J and Toriumi A 2002 *IEEE T. Electron Dev*. **49** 1392
[4] Conibeer G *et al.* 2006 *Thin Solid Films* **511-512** 654
[5] Maeda Y, Tsukamoto N, Yazawaa Y, Kanemitsu Y and Masumoto Y 1991 *Appl. Phys. Lett.* **59** 3168
[6] Choi W K, Chim W K, Heng C L, Teo L W, Ho V, Ng V, Antoniadis D A and Fitzgerald E A 2002 *Appl. Phys. Lett*. **80** 2014
[7] Tong S, Liu F, Khitun A, Wang K L and Liu J L 2004 *J. Appl. Phys*. **96** 773
[8] Scarselli M, Masala S, Castrucci P, Crescenzi M D, Gatto E, Venanzi M, Karmous A, Szkutnik P D, Ronda A and Berbezier I 2007 *Appl. Phys. Lett*. **91** 141117
[9] Ortiz M I, Rodriguez A, Sangrador J, Rodriguez T, Avella M, Jimenez J and Ballesteros C 2005 *Nanotechnology* **16** S197
[10] Paine D C, Caragianis C, Kim T T, Shigesato Y and Ishahara T 1993 *Appl. Phys. Lett*. **62**

2842

[11] Shcheglov K V, Yang C M, Vahala K J and Atwater H A 1995 *Appl. Phys. Lett.* **66** 745
[12] Dutta A K 1996 *Appl. Phys. Lett.* **68** 1189
[13] Giri P K, Bhattacharyya S, Kumari S, Das K, Ray S K, Panigrahi B K and Nair K G M 2008 *J. Appl. Phys.* **103** 103534
[14] Kanjilal A, Hansen J L, Gaiduk P, Larsen A N, Cherkashin N, Claverie A, Normand P, Kapelanakis E, Skarlatos D and Tsoukalas D 2003 *Appl. Phys. Lett.* **82** 1212
[15] Chew H G, Zheng F, Choi W K, Chim W K, Foo Y L and Fitzgerald E A 2007 Nanotechnology **18** 065302
[16] Buljan M *et al.* 2010 *Phys. Rev. B* **81** 085321
[17] Zhang B, Shrestha S, Aliberti P, Green M A and Conibeer G 2010 *Thin Solid Films* **518** 5483
[18] Inoue Y, Fujii M, Hayashi S and Yamamoto K 1998 *Solid State Electron.* **42** 1605
[19] Fujii M, Inoue Y, Hayashi S and Yamamoto K 1996 *Appl. Phys. Lett.* **68** 3749
[20] Fujii M, Mamezaki O, Hayashi S and Yamamoto K 1998 *J. Appl. Phys.* **83** 1507
[21] Yu D, Wang C, Wehrenberg B L and Guyot-Sionnest P 2004 *Phys. Rev. Lett.* **92** 216802
[22] Zhao J, Rebohle L, Gebel T, Von Borany J and Skorupa W 2002 *Solid State Electron.* **46** 661
[23] Zhang B, Shrestha S, Green M A and Conibeer G 2010 *Appl. Phys. Lett.* **96** 261901
[24] Reeves G K and Harrison H B 1982 *IEEE Electron Dev. Lett.* **3** 111
[25] Norris D J, Efros A L and Erwin S C *Science* **319** 1776
[26] Björk M T, Schmid H, Knoch J, Riel H and Riess W 2009 *Nat. Nanotechnol.* **4** 103
[27] Lannoo M, Delerue C and Allan G 1995 *Phys. Rev. Lett.* **74** 3415
[28] Melnikov D V and Chelikowsky J R 2004 *Phys. Rev. B* **69** 113305
[29] Kingston R H 1956 *J. Appl. Phys.* **27** 101
[30] Tsipas P and Dimoulas A 2009 *Appl. Phys. Lett.* **94** 012114
[31] Dimoulas A, Tsipas P, Sotiropoulos A and Evangelou E K 2006 *Appl. Phys. Lett.* **89** 252110
[32] Hanrath T and Korgel B A 2005 *J. Phys. Chem. B* **109** 5518
[33] Zhang S X, Hemesath E R, Perea D E, Wijaya E, Lensch-Falk J L and Lauhon L J 2009 *Nano. Lett.* **9** 3268
[34] Park J, Ryu B, Moon C and Chang K J 2010 *Nano. Lett.* **10** 116
[35] Sharp I D *et al.* 2005 *J. Appl. Phys.* **97** 124316
[36] Djurabekova F and Nordlund K 2008 *Phys. Rev. B* **77** 115325
[37] Hanrath T and Korgel B A 2004 *J. Am. Chem. Soc.* **126** 15466
[38] Molle A, Bhuiyan M N K, Tallarida G and Fanciulli M 2006 *Mater. Sci. Semicond. Process.* **9** 673
[39] Popova L I, Atanassova E D, Kolev D I and Nikolova B M 1986 *J. Non-Cryst. Solids* **85** 382
[40] Nakamura M, Mochizuki Y, Usami K, Itoh Y and Nozaki T 1984 *Solid State Commun.* **50** 1079
[41] Pereira R N, Skov Jensen J, Chevallier J, Bech Nielsen B and Nylandsted Larsen A 2007 *J. Appl. Phys.* **102** 044309

[42] Claeys C and Simoen E 2007 *Germanium-Based Technologies: From Materials to Devices* (Oxford: Elsevier BV) p 47
[43] Greytak A B, Lauhon L J, Gudiksen M S and Lieber C M 2004 *Appl. Phys. Lett.* **84** 4176
[44] Tutuc E, Chu J O, Ott J A and Guha S 2006 *Appl. Phys. Lett.* **89** 263101
[45] Perea D E, Hemesath E R, Schwalbach E J, Lensch-Falk J L, Voorhees P W and Lauhon L J 2009 Nat. Nanotechnol. **4** 315
[46] Melnikov D V and Chelikowsky J R 2004 *Phys. Rev. Lett.* **92** 046802
[47] Hao X J, Cho E-C, Flynn C, Shen Y S, Park S C, Conibeer G and Green M A 2009 Sol. Energ. Mat. & Sol. Cel. **93** 273
[48] Dingle R, Störmer H L, Gossard A C and Wiegmann W 1978 *Appl. Phys. Lett.* **33** 665
[49] Kerr M J, Cuevas A and Sinton R A 2002 *J. Appl. Phys.* **91** 399
[50] Terry M L, Straub A, Inns D, Song D and Aberle A G 2005 *Appl. Phys. Lett.* **86** 172108
[51] Park S, Cho E, Song D, Conibeer G and Green M A 2009 *Sol. Energy Mater. Sol. Cells* **93** 684
[52] Hong S H, Park J H, Shin D H, Kim C O, Choi S H, Kim K J 2010 *Appl. Phys. Lett.* **97** 072108

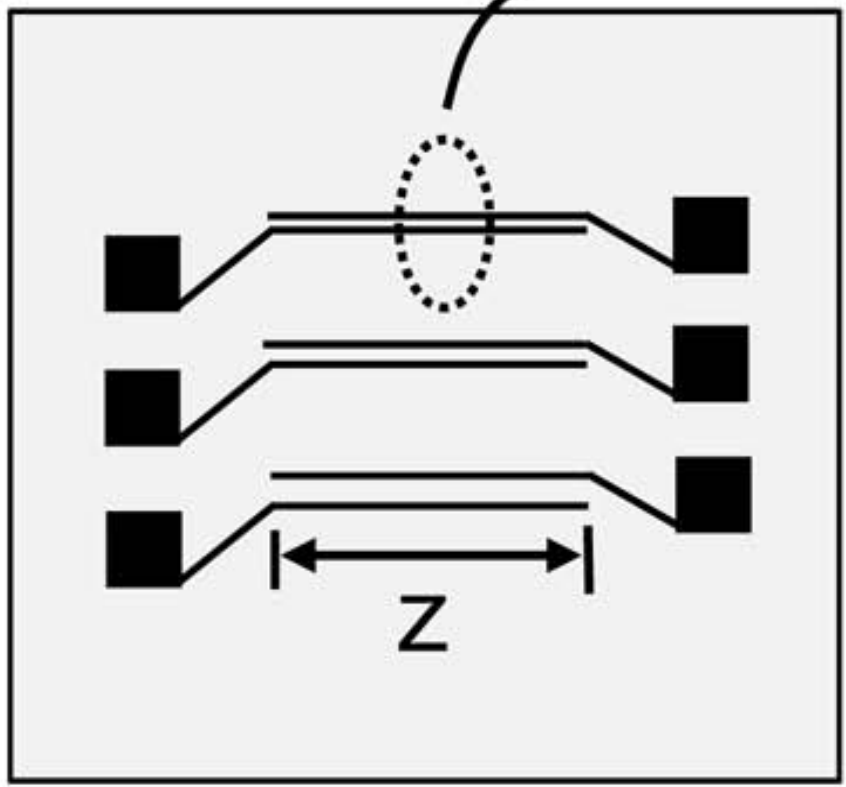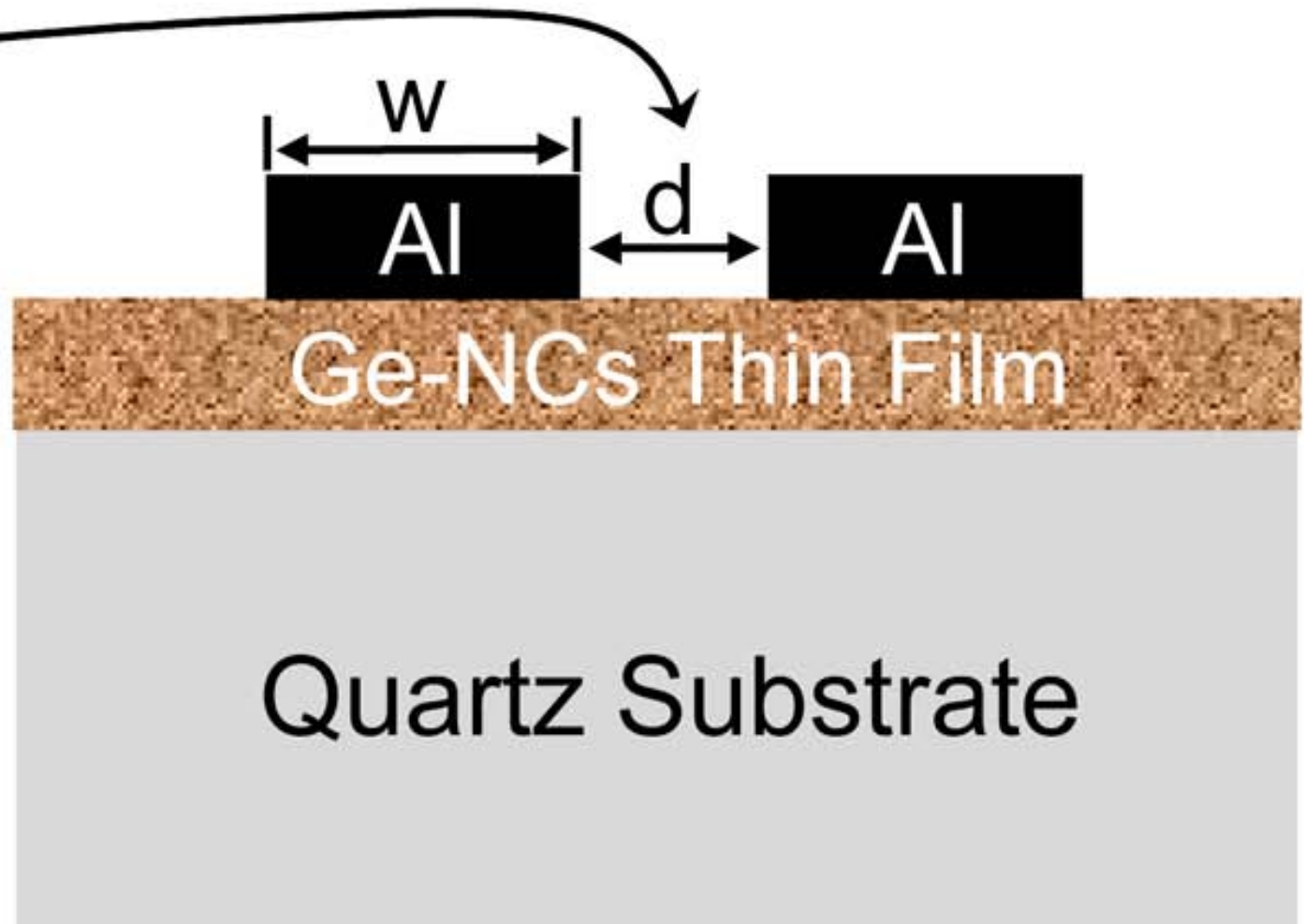

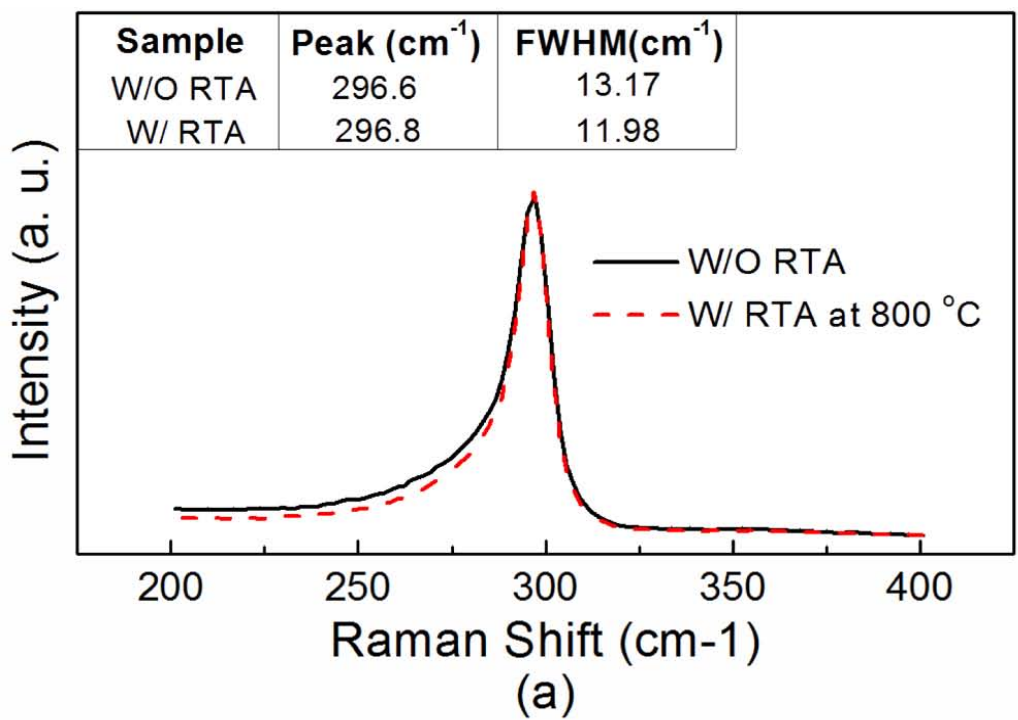

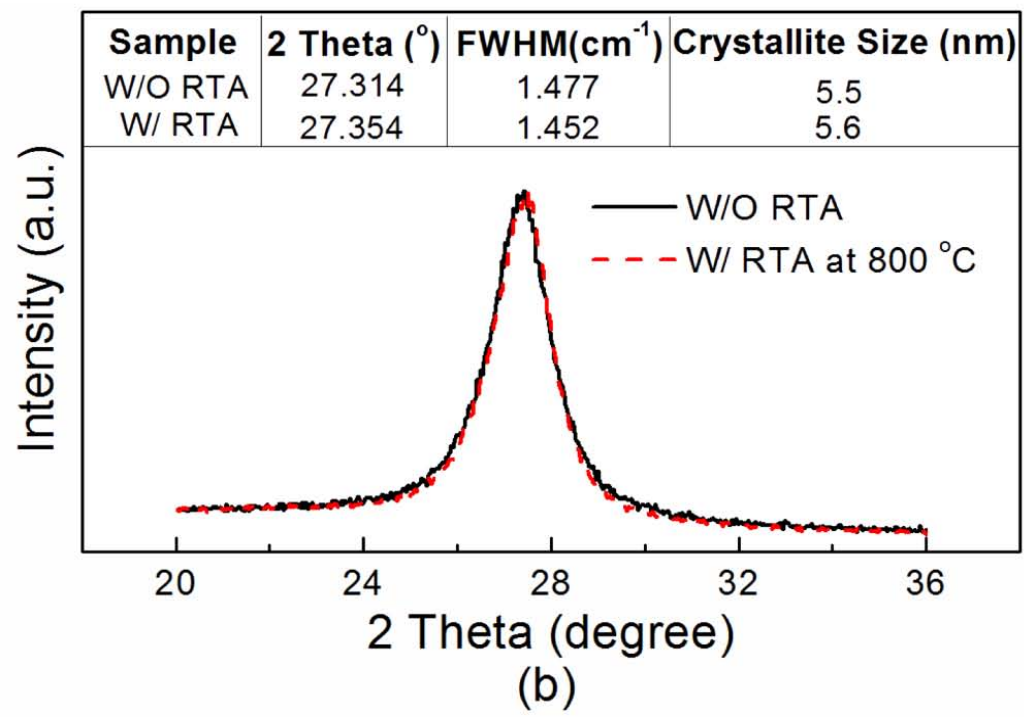

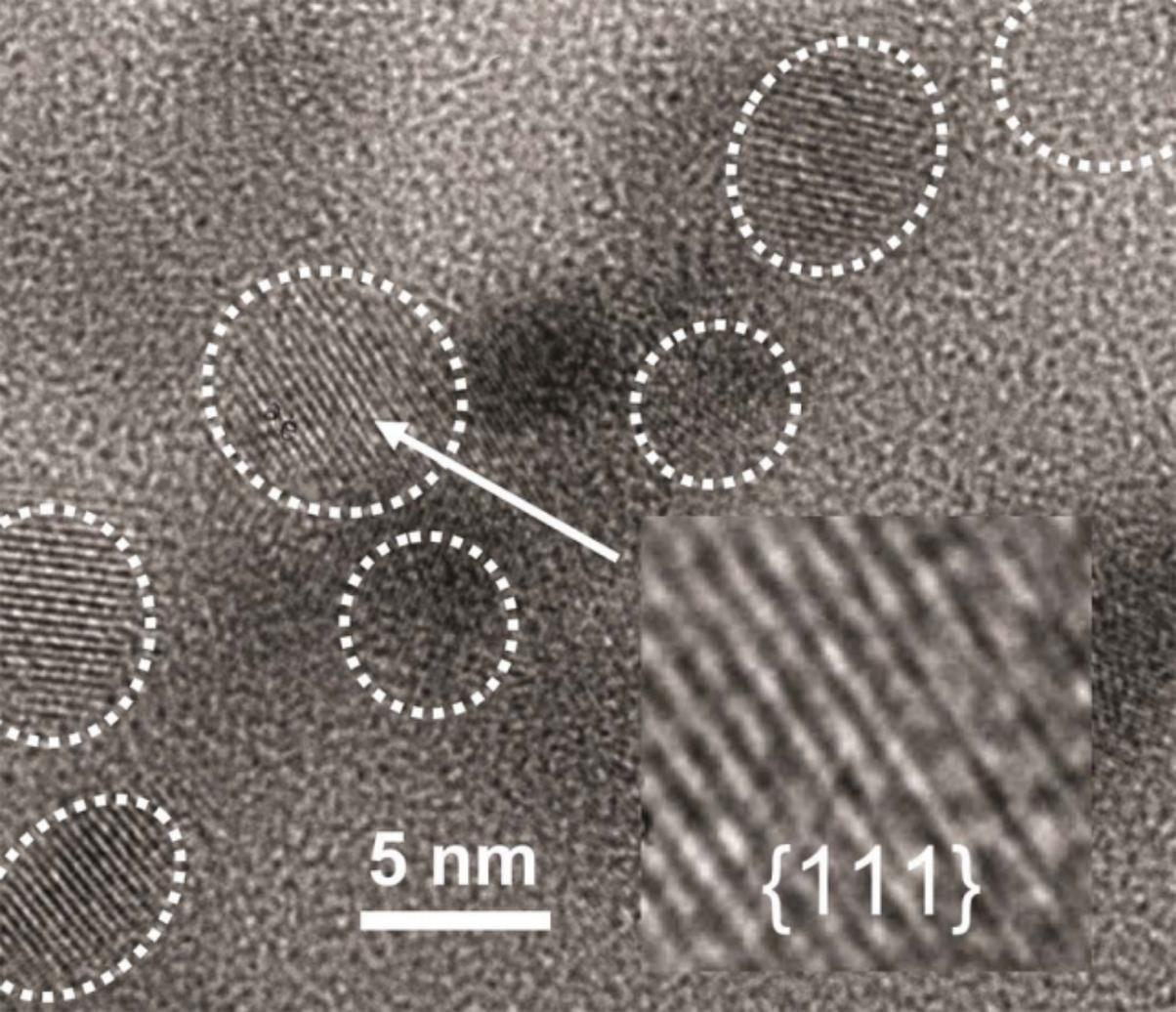

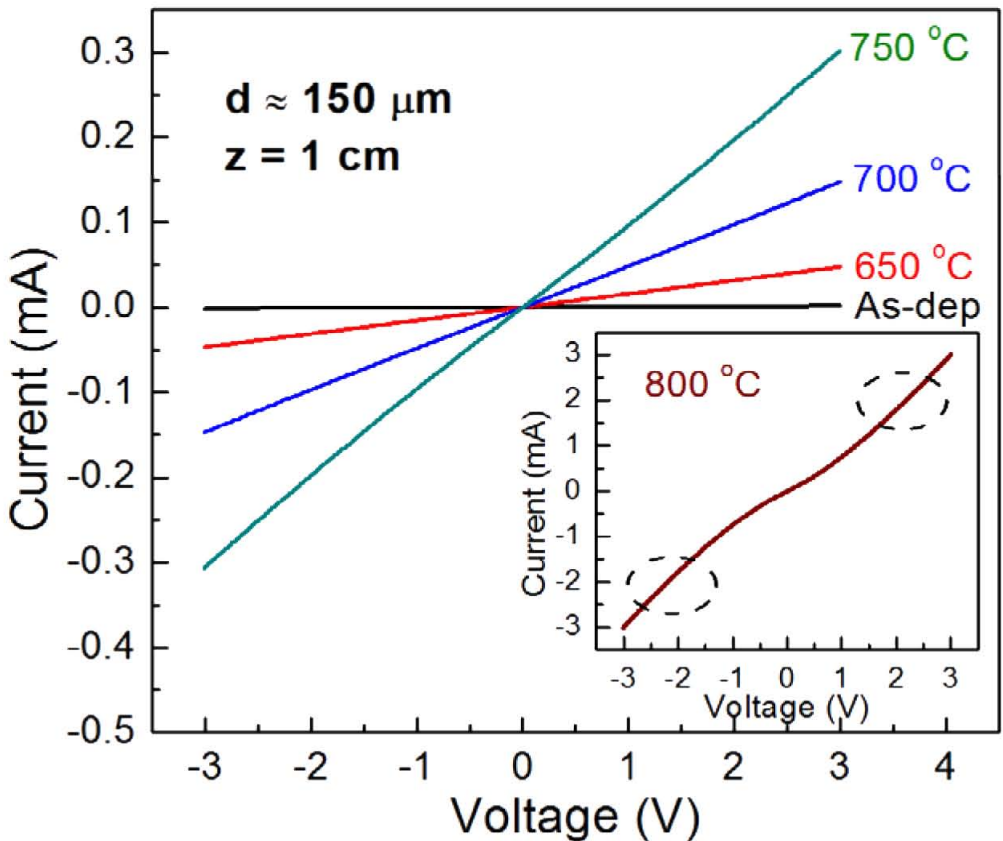

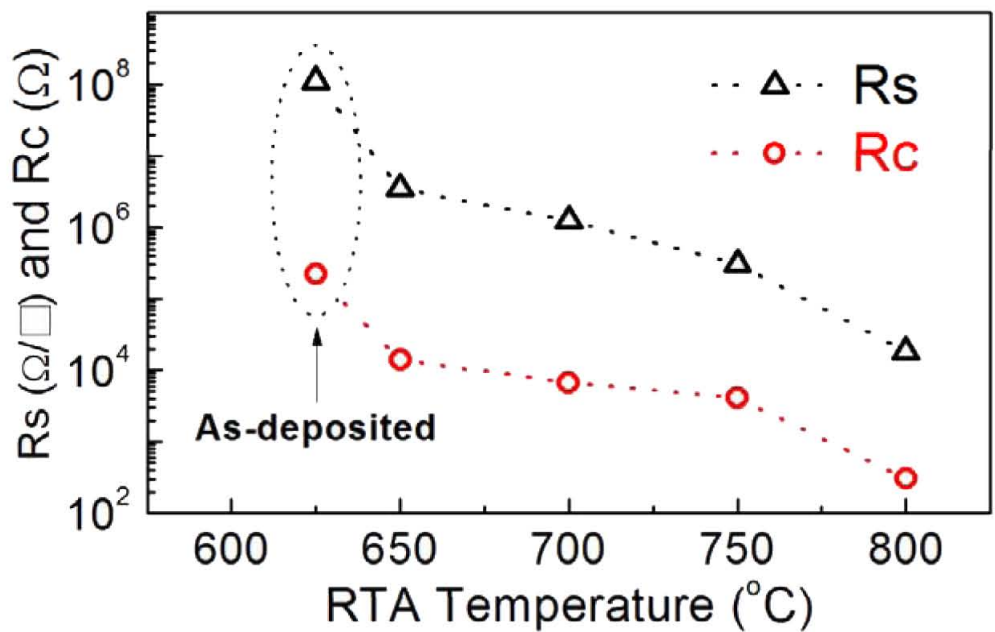

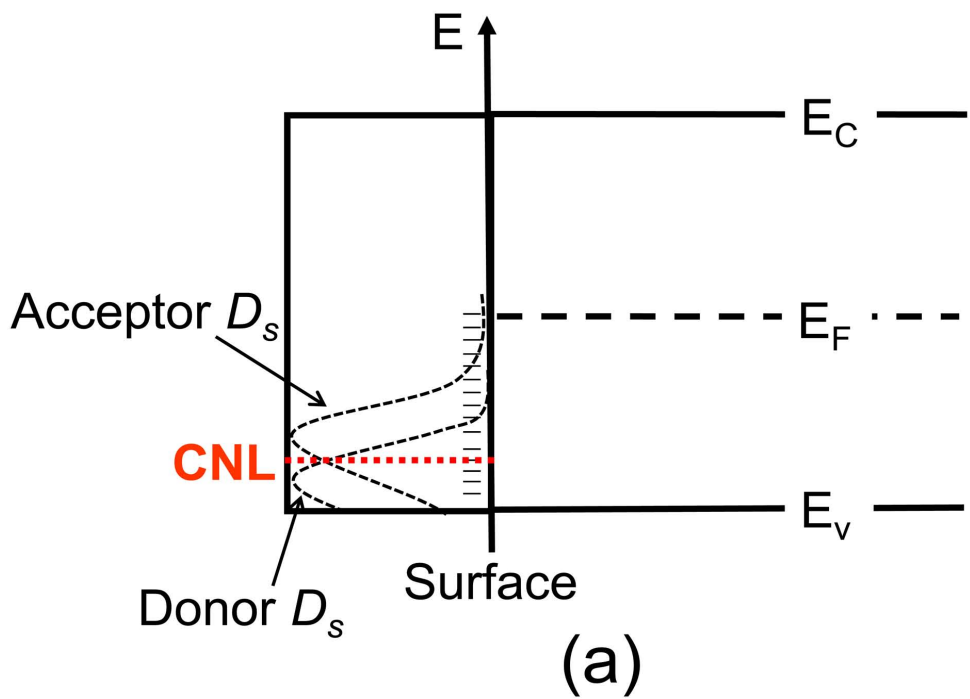

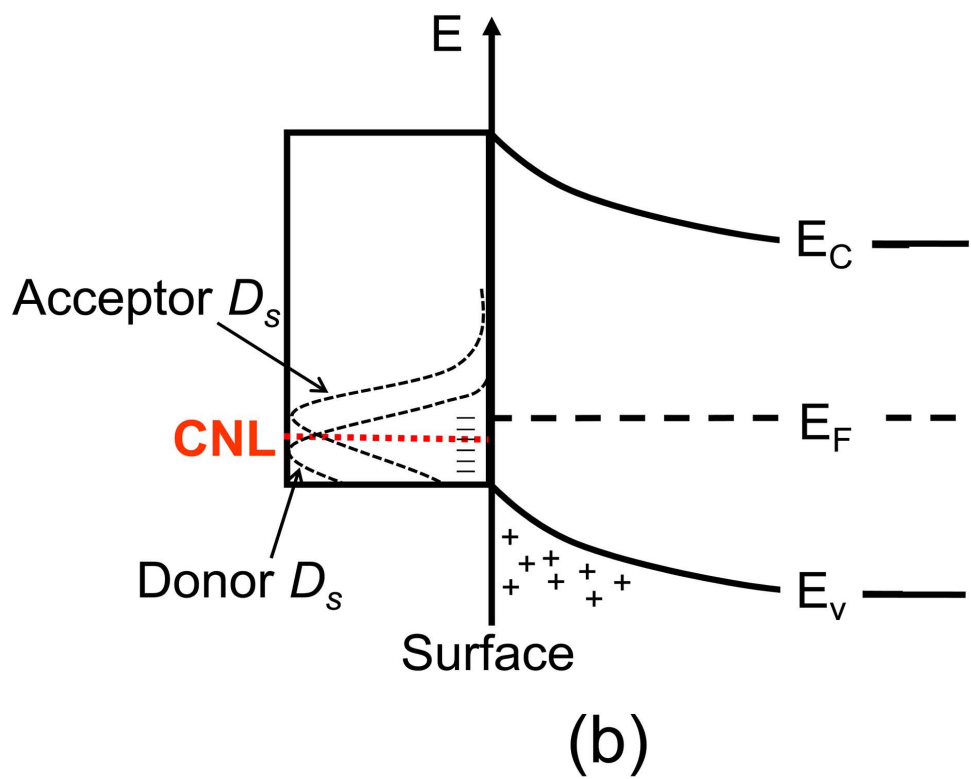

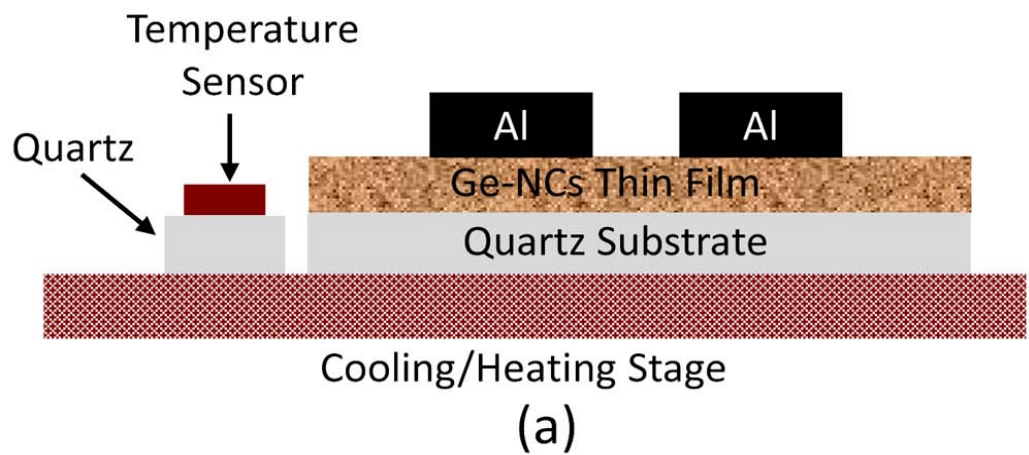

(a)

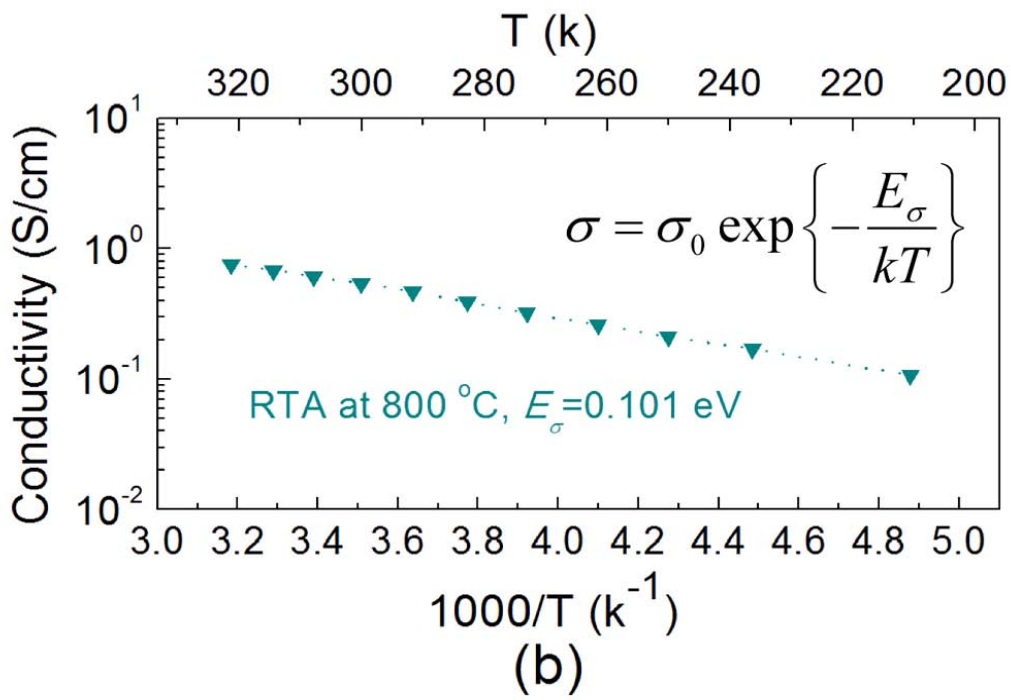

(b)

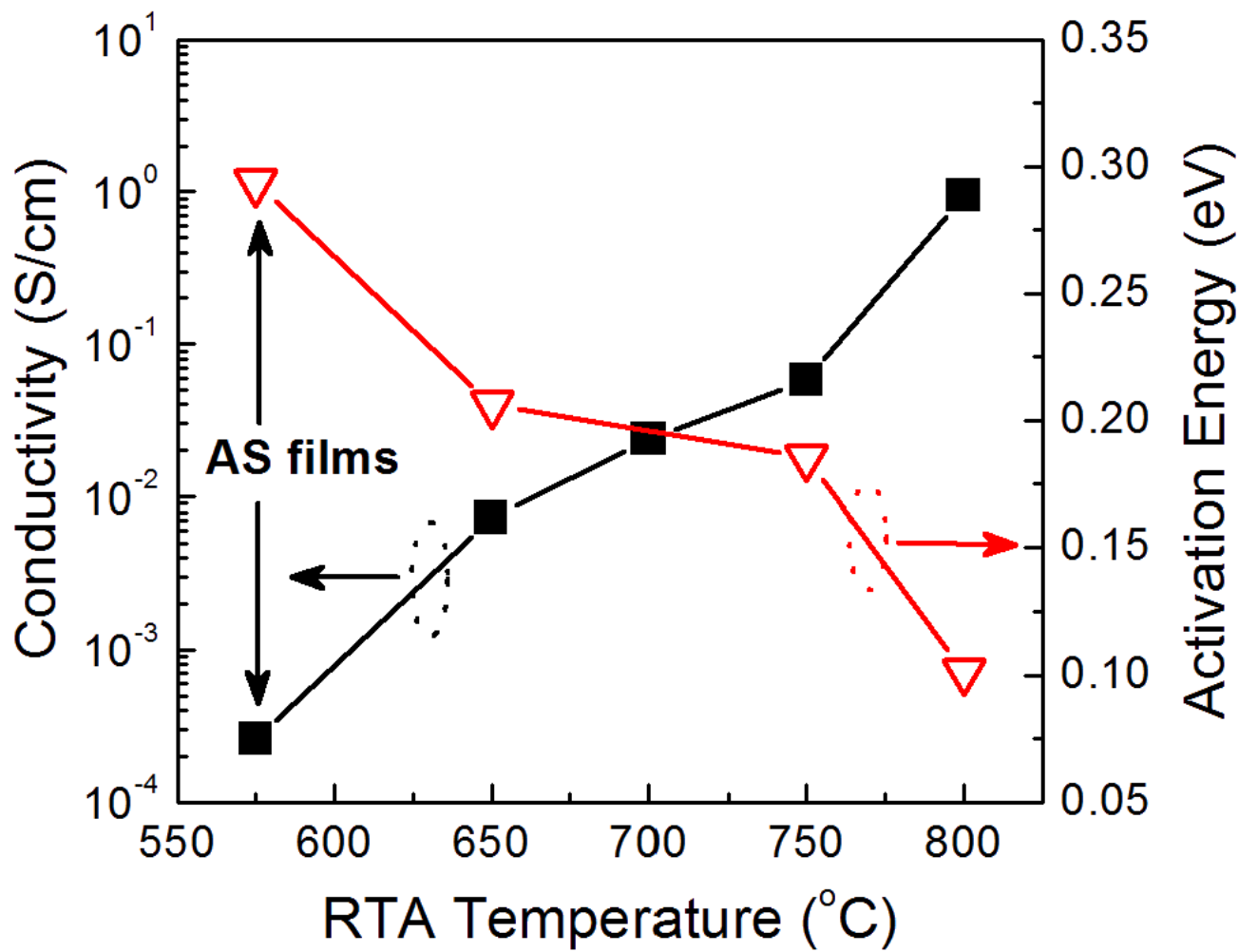

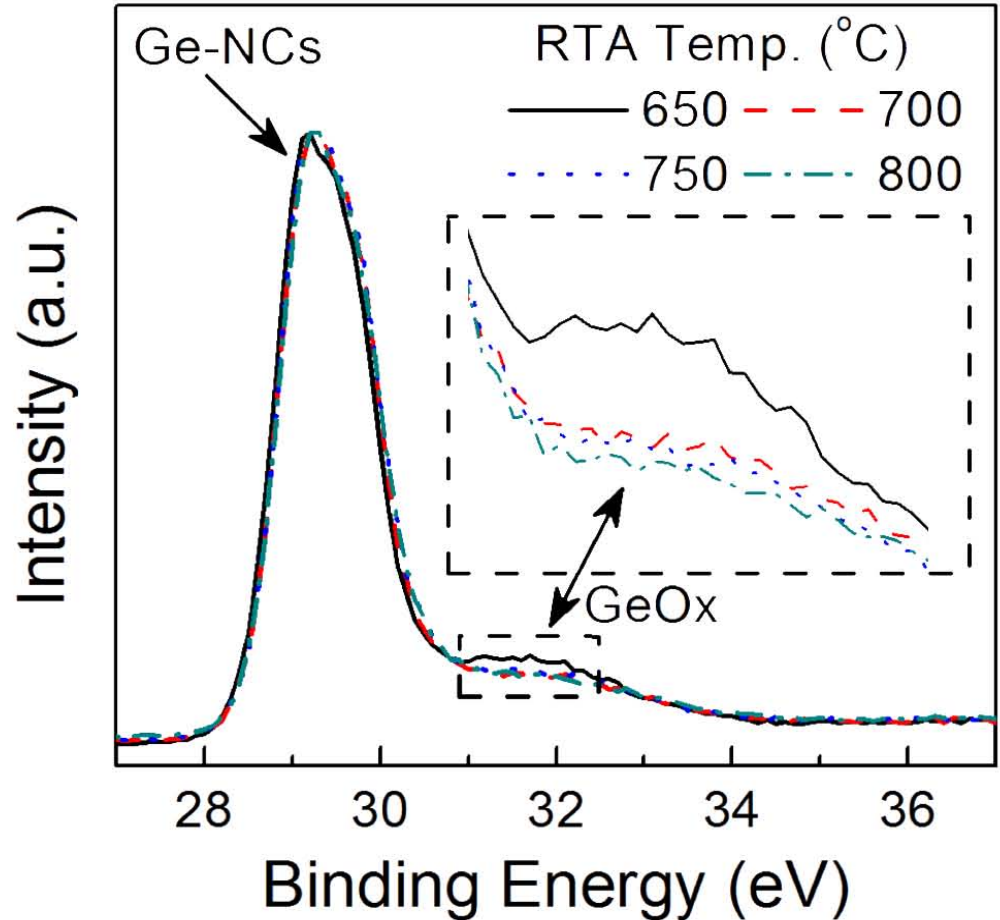

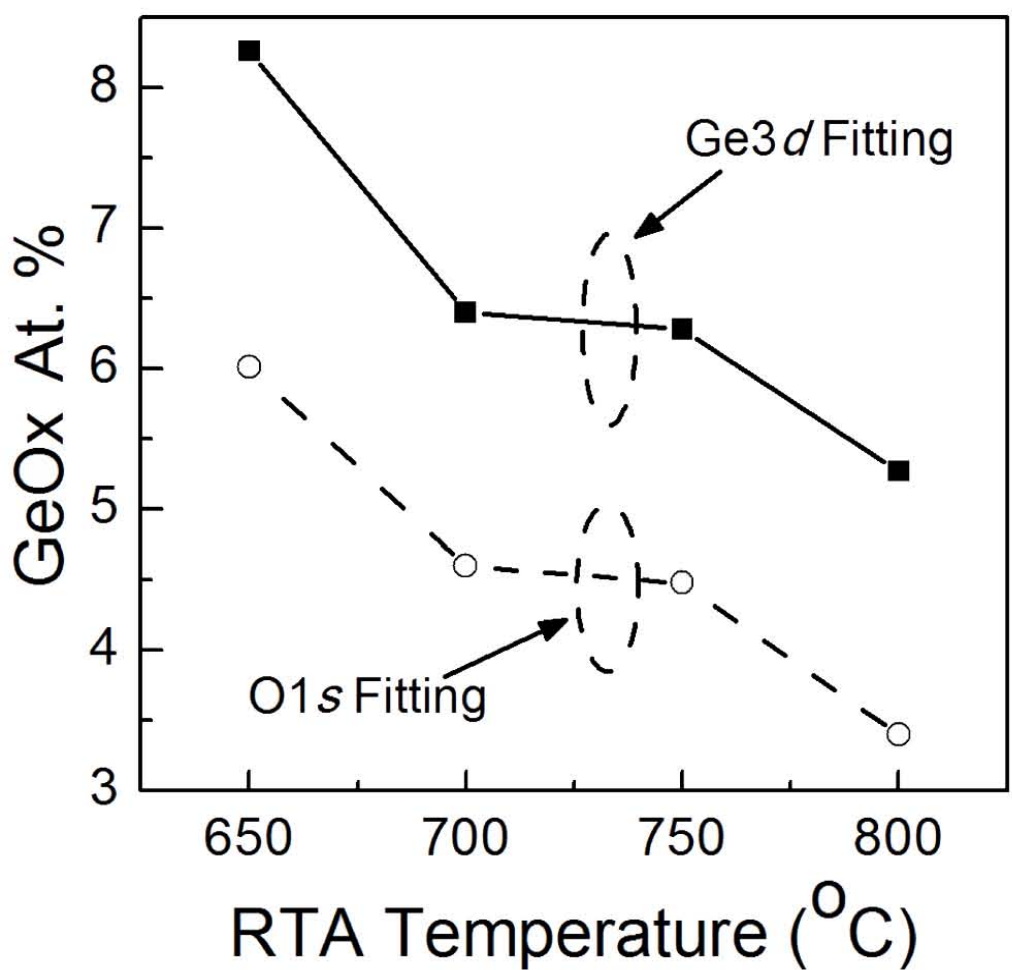

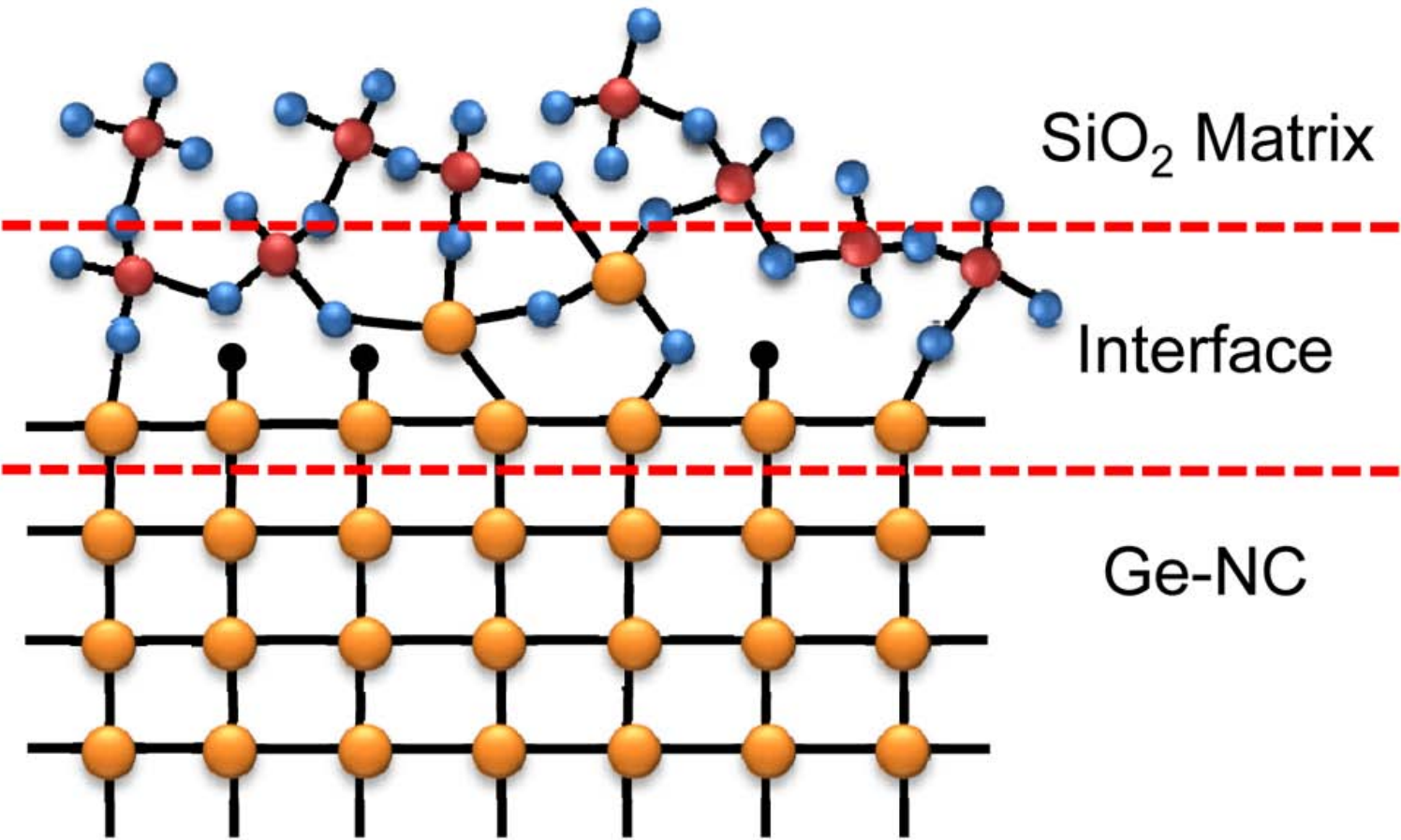

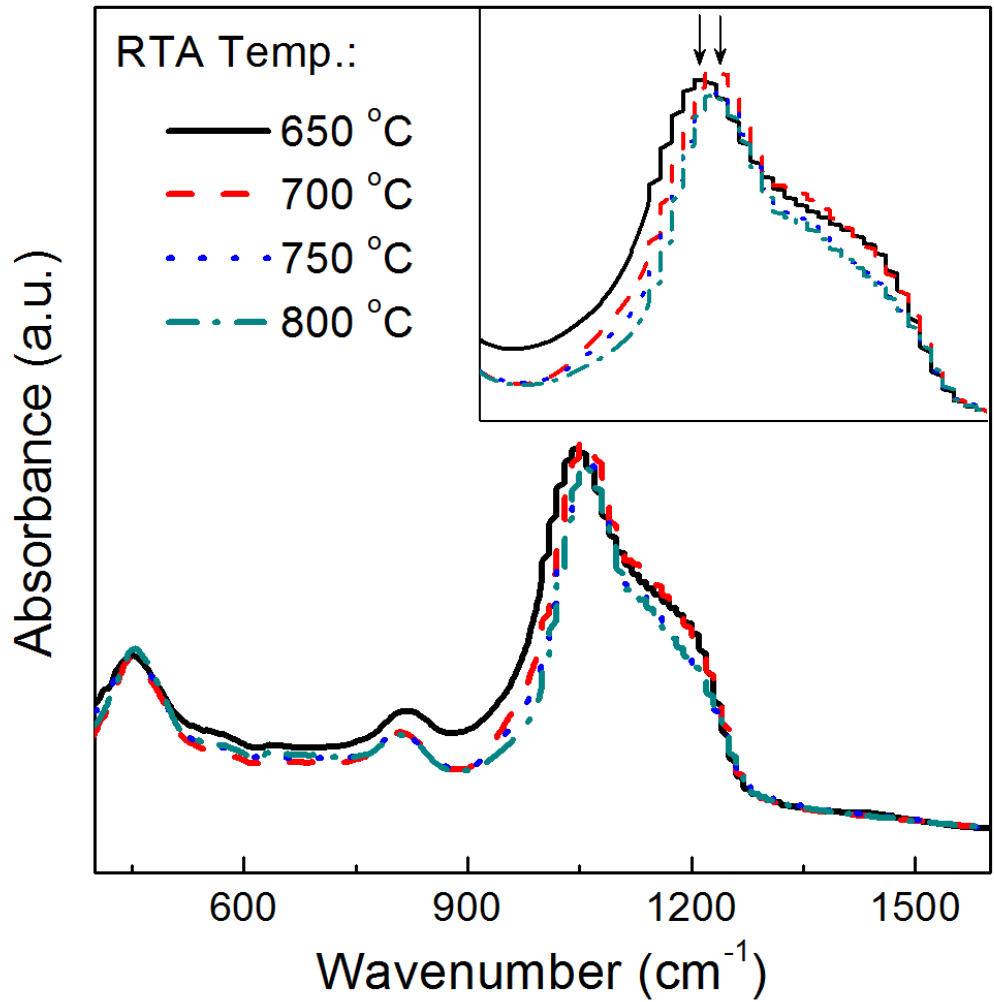

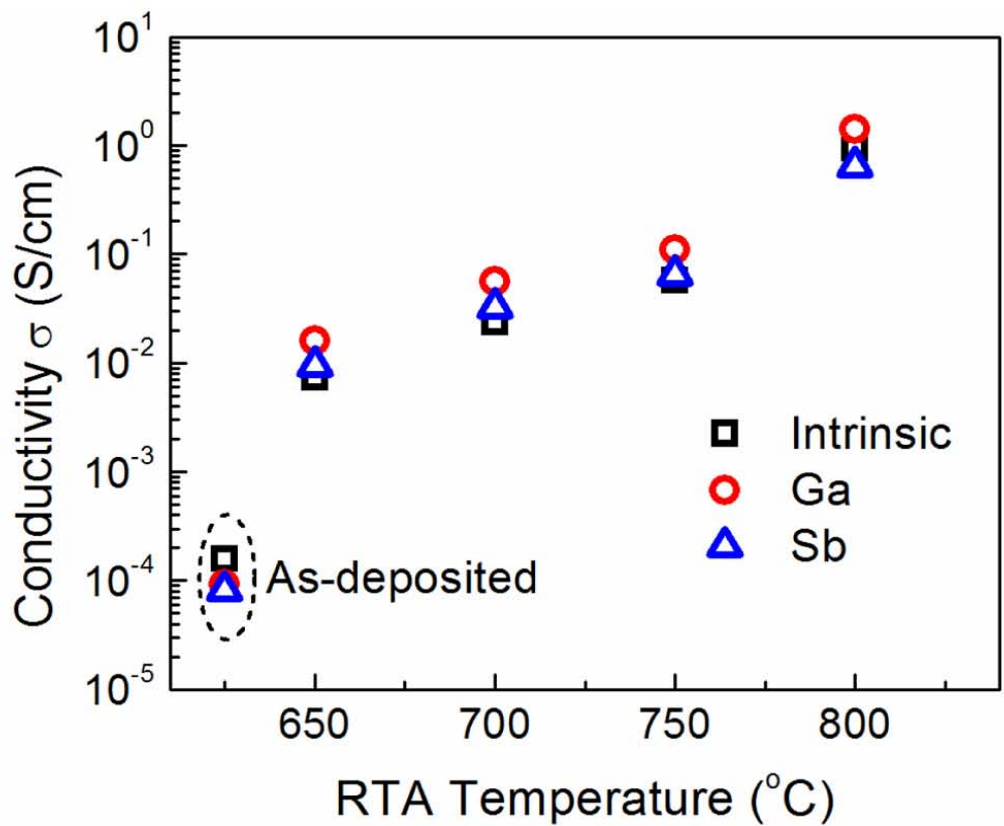

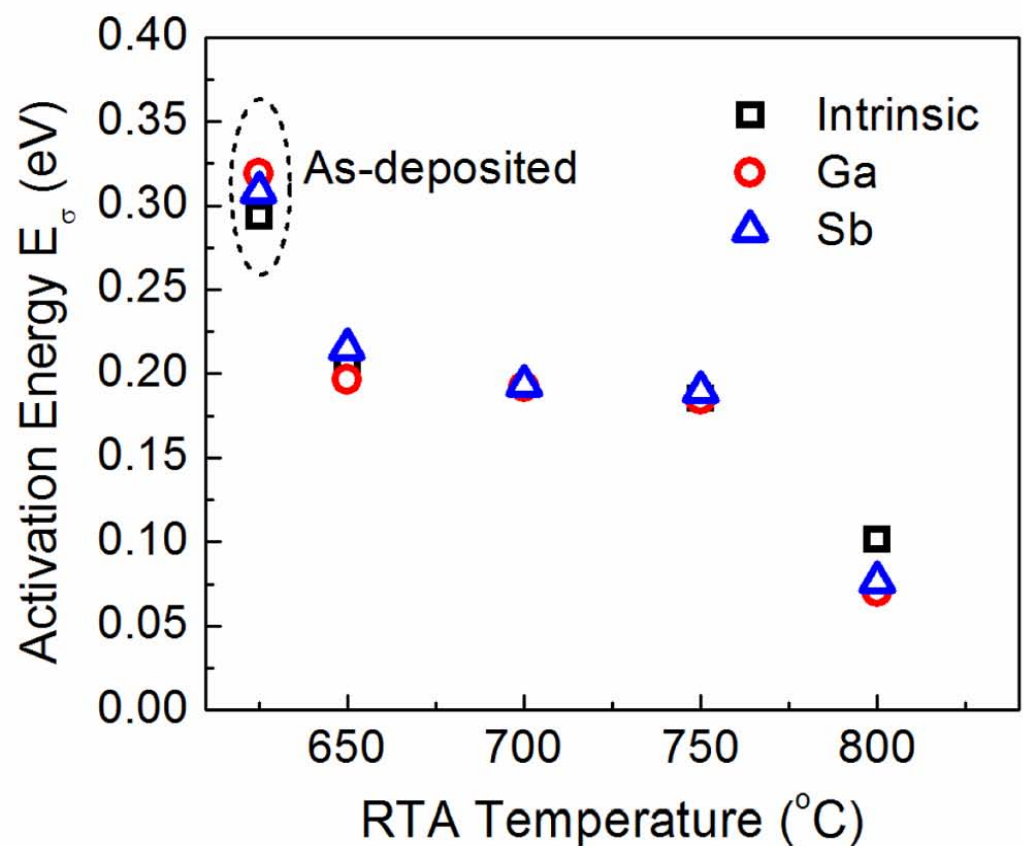

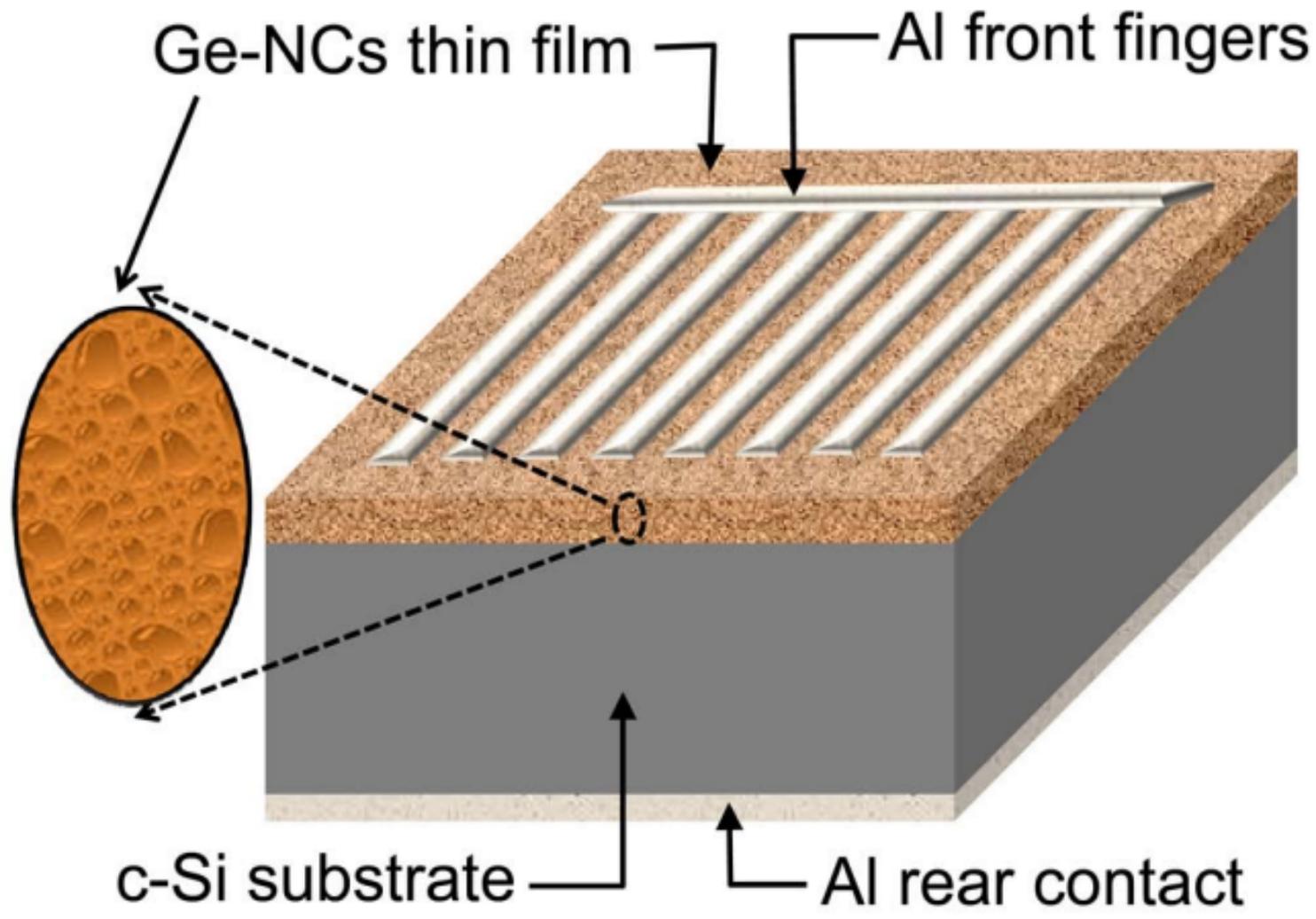

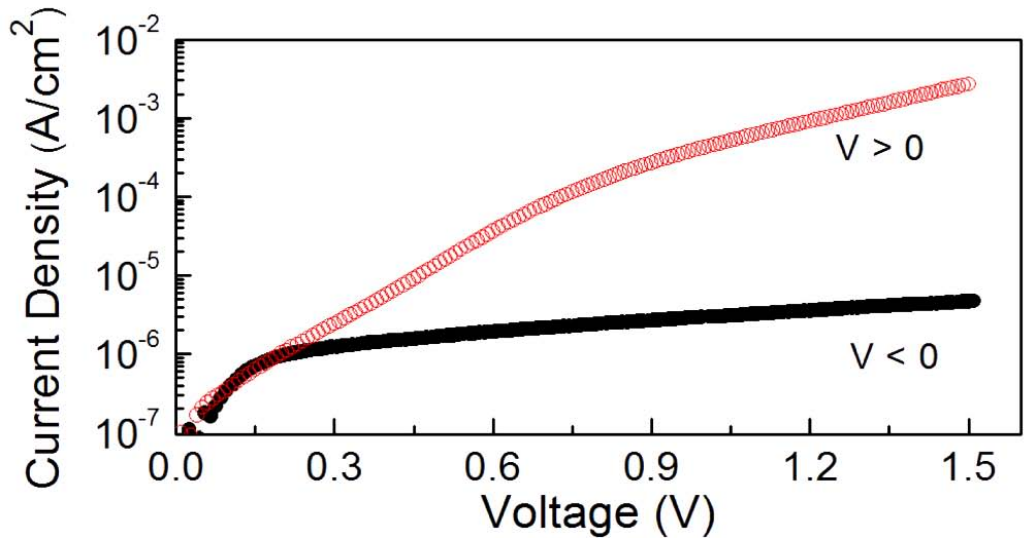

(a)

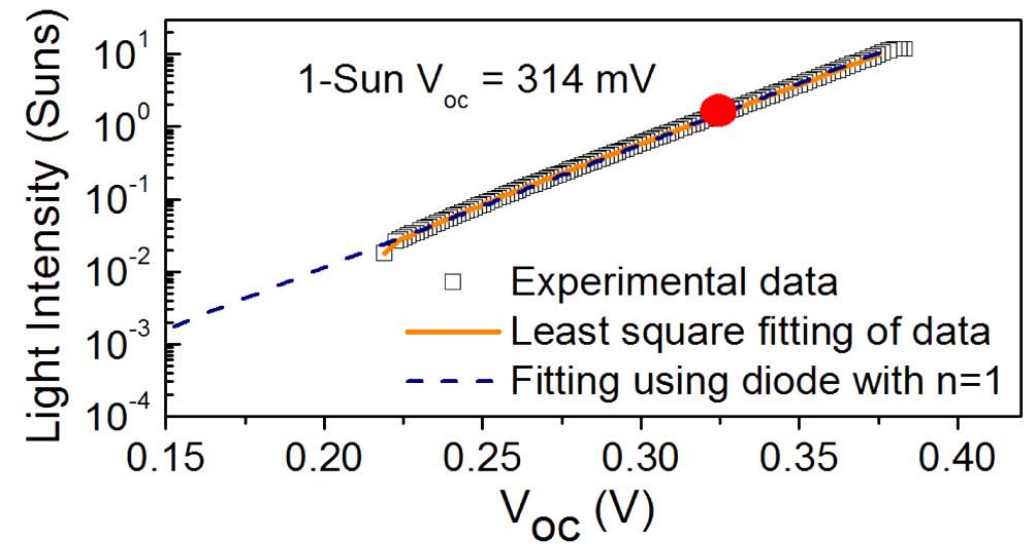

1-Sun $V_{oc}$ = 314 mV

□ Experimental data
— Least square fitting of data
-- Fitting using diode with n=1

(b)